\documentclass[twocolumn,trackchanges]{aastex61}
\pdfoutput=1 
\usepackage{amsmath,amstext}
\usepackage[T1]{fontenc}
\usepackage{apjfonts} 
\usepackage[figure,figure*]{hypcap}
\usepackage{ulem} 
\usepackage{comment}

\newcommand{\fracbrac}[2]{\left(\frac{#1}{#2}\right)}
\newcommand{\ecross}{e_\text{cross}}
\newcommand{\eX}{e_\text{cross}}
\newcommand{\ZZ}{{\cal Z}}

\shorttitle{Resonance Overlap}
\shortauthors{Hadden \& Lithwick}

\begin{document}

\title{A Criterion for the Onset of Chaos in Systems of Two Eccentric Planets}
\author{Sam~Hadden}
\affiliation{Harvard-Smithsonian Center for Astrophysics, 60 Garden St., MS 51, Cambridge, MA 02138, USA}
\correspondingauthor{Sam~Hadden}
\email{samuel.hadden@cfa.harvard.edu}
\author{Yoram Lithwick}
\affiliation{Department of Physics \& Astronomy, Northwestern University, Evanston, IL 60208, USA}
\affiliation{Center for Interdisciplinary Exploration and Research in Astrophysics (CIERA), USA}
\begin{abstract}
We derive a criterion for the onset of chaos in systems consisting of two massive, eccentric, coplanar planets.
Given the planets' masses and separation, the criterion predicts the critical eccentricity above which 
chaos is triggered.  Chaos occurs where mean motion resonances overlap, as in
\citet{Wisdom80}'s pioneering work.  {But whereas Wisdom considered the overlap of first-order resonances only, limiting the applicability of his criterion to nearly circular planets,} we extend  his results to {arbitrarily} eccentric planets (up to crossing orbits) by examining
resonances of all orders. 
We thereby arrive at a simple expression for the critical eccentricity. 
We do this first for a test particle in the presence of a planet, and then generalize to the case of
two massive planets, based on a new approximation to the Hamiltonian \citep{Hadden_inprep}. 
We then confirm our results with detailed numerical simulations. 
Finally, we explore the extent to which  chaotic  two-planet systems  eventually result in planetary collisions.
\end{abstract}

\section{Introduction}
In his proof of the non-integrability of the restricted three-body problem, \citet{Poincare1899} first identified the possibility of dynamical chaos in the motion of planetary systems.
This result cast doubt on Laplace and  Lagrange's "proof" of the solar system's stability \citep{Laskar2012Stable}.
Eventually, the development of KAM theory  \citep{Kolmogorov:430016,arnold1963small,arnold1963proof,moser1973stable} led to the understanding that the phase spaces of conservative dynamical systems like the $N$-body problem are generally an intricate mix of quasi-periodic and chaotic trajectories.
However, deducing when particular planetary systems are chaotic or not remains an unsolved problem; the rigorous mathematical results of KAM theory are typically of little practical use when applied to realistic astrophysical cases. 
One solution is to turn to numerical simulations:
the determination of the solar system's chaotic nature was finally made possible with the advent of the computers capable of running simulations spanning billions of years {\citep[e.g.,][]{1988Sci...241..433S,1989Natur.338..237L,wisdom_holman1991,SussmanWisdom1992,BatyginLaughlin2008,2009Natur.459..817L}}.
A fairly comprehensive global picture of the regular and chaotic regions of phase-space for test-particle orbits in the solar system has since been established by numerical means \citep{Robutel2001}. 

The discovery of thousands of exoplanetary systems over the past few decades has renewed interest in understanding chaos and dynamical stability in planetary systems.
Most general studies of stability in multi-planet systems have focused on fitting empirical relations to large ensembles of $N$-body simulations \citep[e.g.,][]{1996Icar..119..261C,2007MNRAS.382.1823F,Smith:2009eu,Petrovich2015,PuWu2015,2016ApJ...832L..22T,Obertas2017}.
However, such numerical studies suffer some limitations: the large parameter space of the problem, six dynamical degrees of freedom plus a mass for each planet, severely restricts the extent of any numerical explorations.
Additionally, the ages of many exoplanet systems, as measured in planet orbital periods, are frequently orders of magnitude larger than what can feasibly be integrated on a computer so that it is often necessary to extrapolate such numerical results.
Perhaps most importantly, empirical fits do not reveal the underlying dynamical mechanisms responsible for chaos and instability.
Therefore, analytic results are desirable as a complement to such numerical studies.

The resonance overlap criterion, proposed by  \citet{1979PhR....52..263C} \citep[and also ][]{WalkerFord69}, provides one of the few analytic tools for predicting chaos in conservative systems. 
The heuristic criterion states that large-scale chaos arises in the phase space of conservative systems when domains of resonant motion overlap with one another. 
The criterion was first applied to celestial mechanics by \citet{Wisdom80}, who derived a criterion for the onset of chaotic motion of a closely spaced test particle in the restricted circular three-body problem.
Wisdom's criterion is based on the overlap of {\it first-order} mean motion resonances (MMRs).
Since then, the criterion has found numerous applications in planetary dynamics \citep[e.g.,][]{Holman:1996tu,1997AJ....114.1246M,Murray:1999ff,2006ApJ...639..423M,Quillen:2006ez,2008LNP...760...59M,2011ApJ...739...31L,2011MNRAS.418.1043Q,QuillenFrench2014,2015ApJ...799..120B,Ramos:2015vla,StorchLai2015,Petit2017}.
\citet{Wisdom80}'s overlap criterion has been extended to test particles perturbed by an eccentric planet \citep{Quillen:2006ez}  and the case of two massive planets on nearly circular orbits \citep{Deck2013overlap}.
\citet{MustillWyatt2012} derive an analytic criterion for the onset of chaos for an eccentric test particle ($\delta a /a \propto \mu^{1/5} e^{1/5}$), again based on the overlap of first-order MMR's. 

{
The aforementioned works considered the overlap of MMRs only.\footnote{\citet{Ramos:2015vla} refine \citet{Wisdom80}'s overlap criterion by considering the presence of second-order resonances, though they do not account for the finite width of these resonances. {\citet{2008LNP...760...59M} develops a criterion for the overlap of $N$:1 resonances in the general three-body problem to predict chaos in eccentric systems in the widely spaced regime (period ratios $P'/P>2$), complementary to the closely spaced regime we consider in this paper.}}  
As \citet{Wisdom80} originally demonstrated, the widths of first-order resonances increase with increasing eccentricity so that resonance overlap and chaos is expected to occur at wider spacings for eccentric planets than for nearly circular planets.
However, as we demonstrate in this paper,  accounting for the contribution of higher-order resonances beyond first order is essential for correctly predicting the onset of chaos when planets have nonnegligible eccentricities.}

This paper is organized as follows.
We analytically predict {the onset of} chaos based on the overlap of resonances in Section \ref{sec:two_planet_overlap} and compare analytic predictions with numerical integrations in Section \ref{sec:numerical_compare}. 
We compare the newly derived resonance overlap criterion with other stability criteria in \ref{sec:criteria_compare} and {numerically explore} the relationship between chaos and instability in \ref{sec:chaos_vs_stability}. We conclude in Section \ref{sec:conclusion}.

\section{A Theory for the onset of chaos}
\label{sec:two_planet_overlap}
Here we derive the main result of this paper: the resonance overlap criterion that predicts
the critical eccentricity for the onset of chaos, as a 
   function of planet mass and separation.
   To simplify our discussion, we initially restrict our considerations to an eccentric test-particle subject to a massive exterior perturber on a circular orbit (Sections \ref{sec:two_planet_overlap:widths}--   \ref{sec:two_planet_overlap:analytic_expression}). 
  We then generalize  to two planets of arbitrary mass and eccentricity (Section   \ref{sec:two_planet_overlap:generalize}).  This generalization turns out to be surprisingly simple. It is
  based on the discovery by one of us  \citep{Hadden_inprep} of a simple  approximation to the general two-planet Hamiltonian near resonance.

\subsection{Resonance Widths}
\label{sec:two_planet_overlap:widths}

The dynamics of a test-particle near the $j$:$j-k$ MMR  of an exterior circular planet can be approximated by the Hamiltonian  
\begin{align}
    H(&\lambda,\gamma;\delta\Lambda,\Gamma) \approx \delta\Lambda -\frac{3}{4}\delta\Lambda^2 \nonumber \\&+  2\alpha\mu'S_{j,k}(\alpha,e)\cos[(j-k)(t-\lambda) + k\gamma]~, \label{eq:ham_ctp_simple}
\end{align}
which has canonical coordinates $\lambda, \gamma$ and momenta $\delta \Lambda, \Gamma$.  The variables are defined  as follows: $\delta\Lambda=2\left(\sqrt{\frac{a}{a_\text{res}}}-1\right)\approx \frac{a-a_\text{res}}{a_\text{res}}$ where $a$ is the test particle's semimajor axis (and henceforth un-primed orbital elements refer to the test particle); $a_\text{res}$
 is that of  nominal  resonance, i.e., 
 \begin{eqnarray}
 a_\text{res}=\left(\frac{j-k}{j}\right)^{2/3}a' \approx \left(1-{\frac{2k}{3j}} \right) a' \label{eq:mya}
 \end{eqnarray}
  where $a'$ is the planet's semimajor axis, and the latter approximation assumes close spacing; $\lambda$ is the mean longitude; 
  $\Gamma=2\sqrt{\frac{a}{a_\text{res}}}(1-\sqrt{1-e^2})$; 
  $\gamma=-\varpi$, where $\varpi$ is the longitude of perihelion;  $e$ is the eccentricity; $\alpha=a/a'$; $\mu'$ is the ratio of the planet's mass to that of the star; and time units are chosen so that the planet's mean longitude is 
  $\lambda'= {j-k\over j}t$ (or equivalently $\frac{d\lambda}{dt}=1$ when $\delta\Lambda=0$ and $\mu'\rightarrow 0$).  
  The above Hamiltonian is standard \citep[e.g.,][]{2000ssd..book.....M}. 
  {As is common, we approximate the coefficient of the cosine term,  $2\mu'\alpha S_{j,k}(\alpha,e)$, as being temporally constant.  
This ``pendulum'' approximation \citep{2000ssd..book.....M} shows good agreement with exact resonance widths computed via numerical averaging methods \citep[e.g.,][see Appendix \ref{sec:appendix:pendulum_vs_exact} for comparison]{Morbidelli:1995jf} with one notable exception:
 it does not adequately capture the resonant width of $k=1$ resonances at low $e$ {\citep[$\lesssim (\mu'/j)^{1/3}$; see][]{Wisdom80}.} 
 We discuss the consequences of this shortcoming of the pendulum model below.
 }

The Hamiltonian in Equation \eqref{eq:ham_ctp_simple} can now be transformed with the type-2 generating function $F_2= [(j-k)(t-\lambda) + k\gamma]I + \gamma K$ to the new Hamiltonian
\begin{eqnarray}
H'(\phi;I) =-\frac{3}{4}(j-k)^2 I^2 + 2\alpha\mu'S_{j,k}(\alpha,e)\cos\phi~. \label{eq:res_ham_transformed}
\end{eqnarray}
where $I=(k-j)\delta\Lambda$ and $\phi=(j-k)(t-\lambda)+k\gamma$. The Hamiltonian $H'$ describes a pendulum with a maximal libration half-width  
\begin{equation}
\Delta I = \sqrt{\frac{16\alpha \mu'  |S_{j,k}(\alpha,e)|}{3(j-k)^2}} \label{eq:res_width}
\end{equation}
or, in terms of semi-major axis,
\begin{equation}
\frac{\Delta a}{a_\text{res}} = (k-j)\Delta I = \sqrt{\frac{16\alpha\mu'|S_{j,k}(\alpha,e)|}{3}}~.\label{eq:da_ctp}
\end{equation}

\subsection{{The ``Close Approximation'' for $S_{j,k}$}}
\label{sec:close_approx}
The cosine amplitude $S_{j,k}(\alpha,e)$ is often replaced with its leading-order approximation $\propto e^k$, which is valid at low $e$
\citep{2000ssd..book.....M}.  However,  we will consider eccentricities up to
\begin{equation}
e_{\rm cross}\equiv {a'-a\over a} \ ,
\end{equation}
which is the eccentricity at which the particle's orbit crosses the planet's.  The leading-order approximation is inadequate at such high $e$, as we quantify below.
Therefore in Appendix \ref{sec:appendix}, we derive a more accurate approximation  by proceeding as follows:
first, we derive an {\it exact} expression for $S_{j,k}$ in the form of a one-dimensional definite integral (Eq. \ref{eq:Sjk_with_M}).  However, this integral is both cumbersome and numerically challenging to evaluate at high $k$. 
Therefore, we derive a simpler expression under the approximation that the test particle is close to the planet ($a'/a-1 \ll 1$). 
Under this ``close approximation'', 
 the integral simplifies considerably, and furthermore it only depends on $\alpha$, $e$, and $j$ in the combination ${e\over e_{\rm cross}}\approx {3j\over 2k}e$. We thereby find
(Eq. \ref{eq:Sk_def}) 
\begin{multline}
S_{j,k}(\alpha,e)\approx s_k({e\over\eX})
\equiv\\ {1\over\pi^2}
\int_0^{2\pi}K_0\left[\frac{2k}{3}(1+{e\over\eX}\cos M)\right]\cos\left[k\left(M+\frac{4}{3} {e\over\eX}\sin M\right)\right] {dM}
\label{eq:Sjk_approx2} 
\end{multline}
where $K_0$ is a modified Bessel function. 
Equation \eqref{eq:Sjk_approx2} provides an
adequate approximation when planet period ratios are $P'/P\lesssim 2$, generally predicting resonance resonance widths via Equation \eqref{eq:da_ctp} with $\lesssim 20\%$ fractional errors.
\subsection{Resonance Overlap}
\label{sec:two_planet_overlap:optical_depth}
\begin{figure}
\centering
\includegraphics[width=0.45\textwidth]{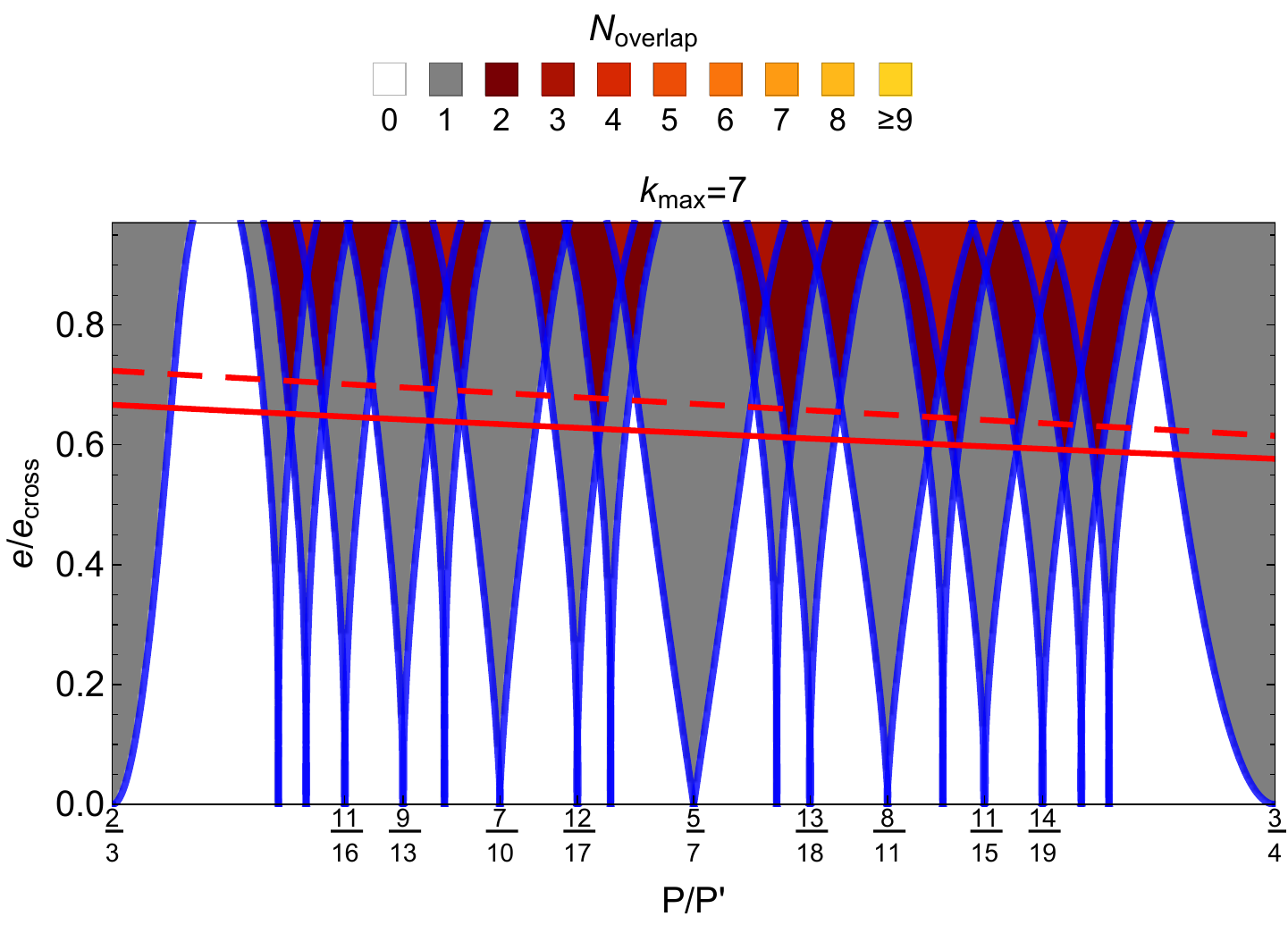}
\includegraphics[width=0.45\textwidth]{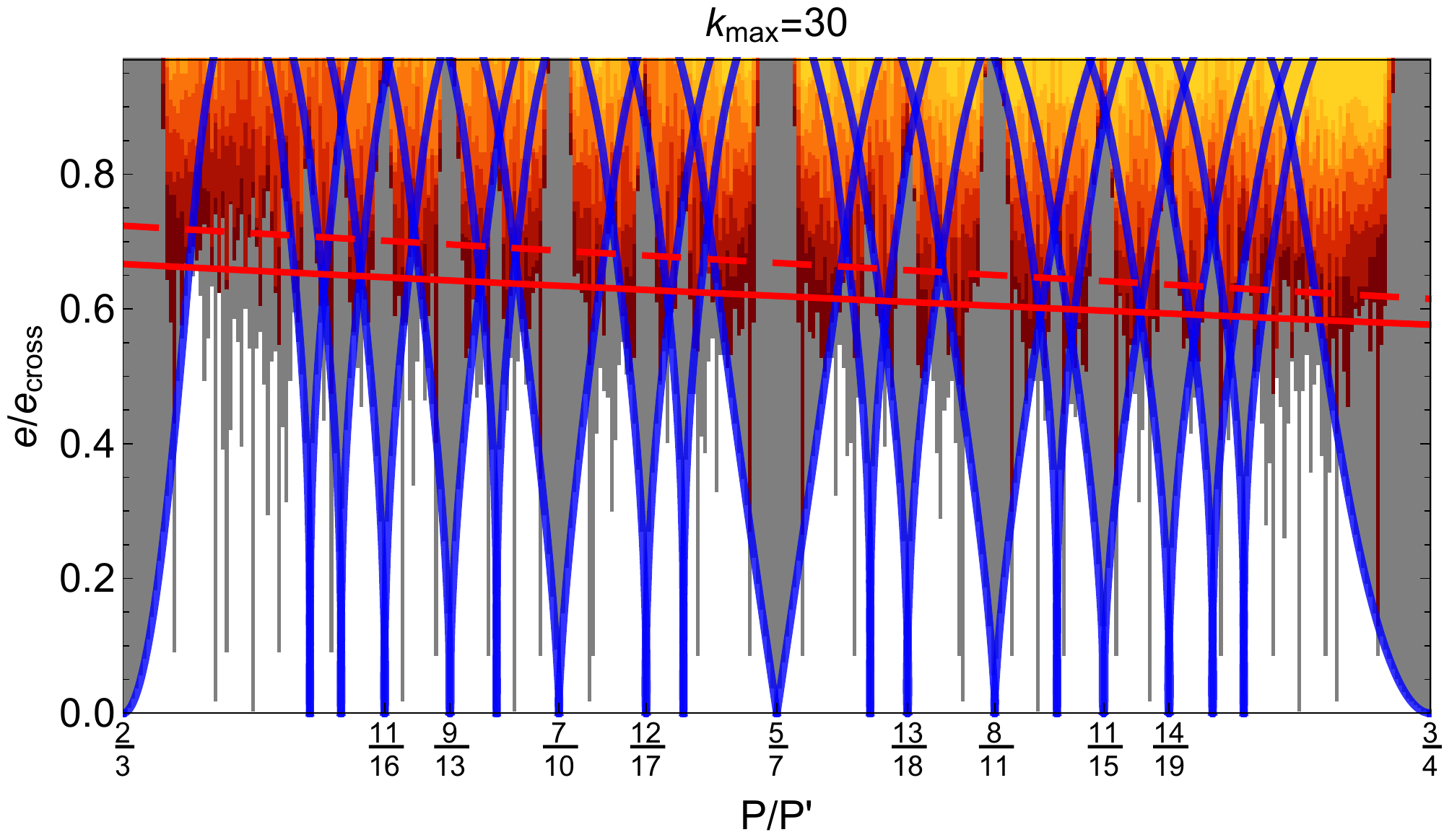}
\caption{\label{fig:optical_depth_schematic}
Structure of resonances and their overlap between the 3:2 and 4:3 MMRs for a test-particle perturbed by a $\mu'=10^{-5}$ planet on a circular orbit.
Blue separatrices are plotted for resonances up  to seventh order. 
Each panel shows a grid of $e/\eX$ versus $P/P'$, the ratio of periods of the test particle to the planet; 
note that $\eX\approx 0.25$ at these periods.
Grid points are colored according to the number of resonances they fall within, accounting for all resonance up to order $k_\text{max}=7$ in the top panel and $k_\text{max}=30$ in the bottom panel.
Resonance widths are computed with $S_{j,k}=s_k$ (Eq. \ref{eq:Sjk_approx2}) for $k>7$ in the bottom panel.
Points falling in $N_\text{overlap}\ge 2$ resonances are predicted to be chaotic based on the resonance overlap criterion.
The solid red line marks the $e/\eX$ value where the covering fraction of resonances is unity, i.e., $\tau_\text{res}=1$ according to Equation \eqref{eq:tau_sum}. 
The dashed red line  marks the estimate of our fitting formula,  Equation \eqref{eq:ecrit_approx_2}, for $\tau_\text{res}=1$.
Note that the widths of the two $k=1$ resonances at the left and right edges of the figure are represented incorrectly near $e=0$ due to our adoption of the pendulum approximation. In reality, they are wider than shown at {$e\lesssim (\mu'/j)^{1/3}$, i.e., at $e/\eX\lesssim 0.05$ in this figure.}}
\end{figure}

Our criterion for chaos is the overlap of resonances \citep{1979PhR....52..263C,Wisdom80}.
With a  formula for resonance widths in hand (Equation \ref{eq:da_ctp}), we examine under what conditions resonances overlap and motion is chaotic. 
The top panel of Figure \ref{fig:optical_depth_schematic} plots the locations and widths for  all resonances with order $k\leq 7$ between the 3:2 and 4:3 MMR's. 
Resonance widths in Figure \ref{fig:optical_depth_schematic} are computed using Equation \eqref{eq:da_ctp} with $S_{j,k}$ computed via Equation \eqref{eq:Sjk_with_M}.
At low $e$, the resonances are narrow, and there is no overlap. As $e$ increases, the resonances widen ($\Delta a\propto e^{k/2}$) and overlap everywhere.
At a given $a$ (or $P/P'$), there is a critical $e$ at which the test particle comes under the influence of two resonances simultaneously and hence becomes chaotic.

Of course, to determine the critical $e$, one should include the overlap
between resonances of all orders, not just $k\leq 7$. 
However, resonances with very high $k$'s will have little effect on the critical $e$ {even though there are an infinite number of them}. That is  because the widths decrease exponentially
with increasing $k$, whereas the number of resonances increases only algebraically. 
The bottom panel of Figure \ref{fig:optical_depth_schematic} illustrates this by repeating the top panel, but with $k\leq30$.  
We see that the regions where resonances are significantly overlapped are similar in the two panels. 

We estimate the critical $e$ for significant overlap by first evaluating the covering fraction (or ``optical depth'' $\tau_{\rm res}$) of resonances in a range $\delta a$
of semimajor axes, as a function of $e$. The threshold for overlap will then be the $e$ at which  $\tau_{\rm res}=1$. 
{(This ``optical depth" construction is similar to \citet{2011MNRAS.418.1043Q}'s method for estimating the density of three-body resonances in systems of three planets.)}
Now, to determine a convenient range $\delta a$, we examine the pattern of non-commensurate MMR's in Figure 
\begin{figure}
\centering
\includegraphics[width=0.45\textwidth]{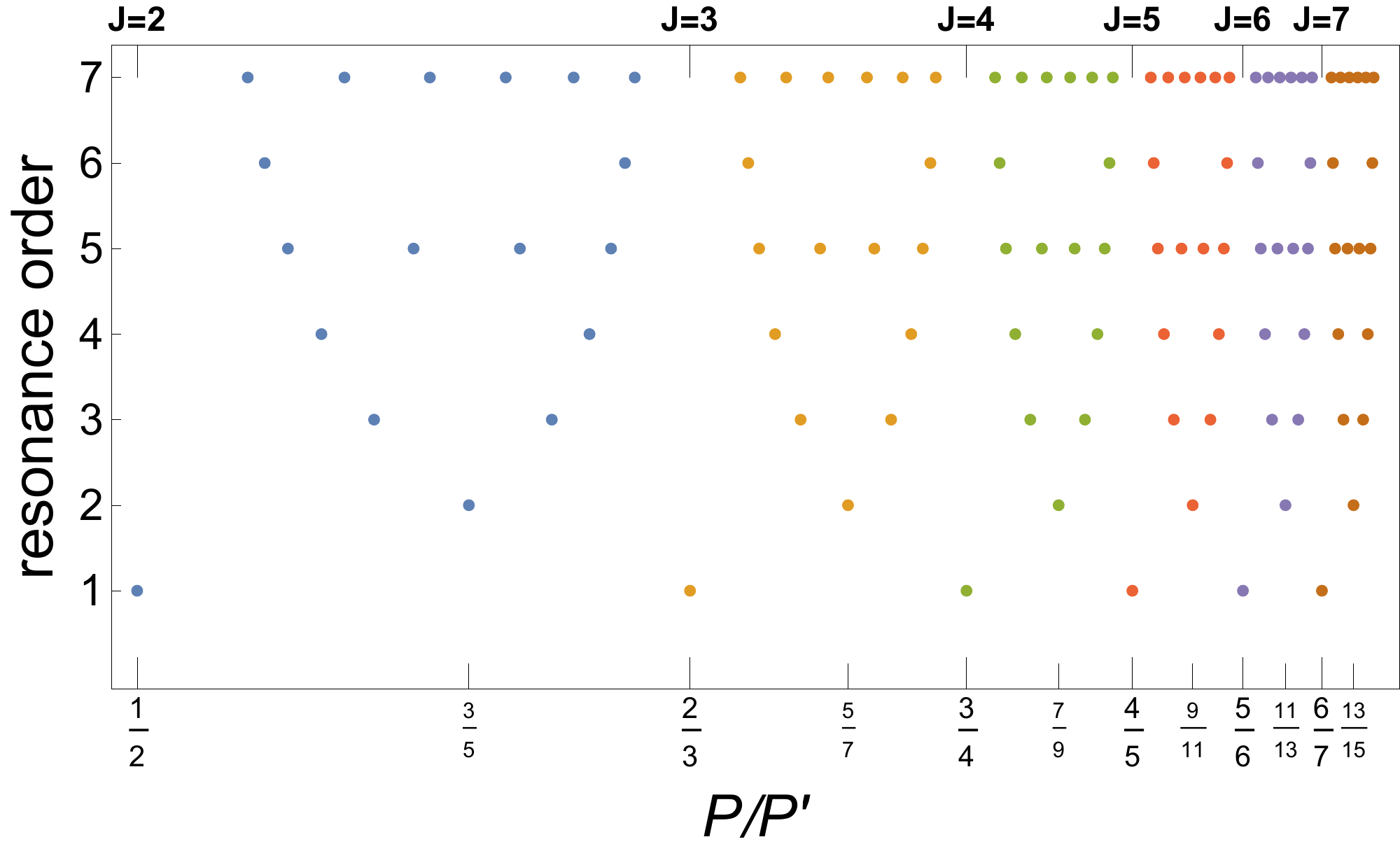}
\caption{\label{fig:res_locations}
The location and order of the  non-commensurate mean motion resonances up to seventh order, illustrating how the pattern repeats itself relative to
each first-order MMR.
Note that each differently colored cluster 
can be enumerated up to any $k$
using the  `Farey sequence' $F_k$, which is the sequence of reduced fractions between 0 and 1 that have denominators less than or equal to $k$, e.g., $F_2=\left\{0,1/2,1\right\}$ and $F_3=\left\{0,1/3,1/2,2/3,1\right\}$.
The resonant period ratios within a cluster that contains the first order  $J:J-1$ MMR
  occur at $P/P' = \frac{J-1+r}{J+r}$ for $r\in F_k$. 
When plotting $P/P'$ elsewhere in this paper we stretch the horizontal scale to assign equal measure to each group of resonances associated with a single $J$.
This is done by setting plots' horizontal coordinates uniformly in $J=(1-P/P')^{-1}$.
}
\end{figure}	 
\ref{fig:res_locations}.
We see that the pattern repeats itself relative to each first-order MMR.  Therefore, we choose $\delta a$
to be the distance between neighboring first-order MMR's, or from Equation \eqref{eq:mya}
\begin{eqnarray}
{\delta a\over a'} \approx {2\over 3J^2}  \approx {3\over 2}\left({a'-a\over a'}  \right)^2
\end{eqnarray}
where $J$ in the above refers to that of the first order $J:J-1$ MMR's. 
To evaluate the covering fraction of resonances within this semi-major axis range, we assume that the planet and particle are sufficiently close  that we can treat $\eX$ as constant over the range.
Taking the resonant width $\Delta a$ from Equation \eqref{eq:da_ctp} and using the close approximation of Equation (\ref{eq:Sjk_approx2}) yields
 \begin{eqnarray}
\tau_\text{res}&\approx&
{1\over\delta a }\sum_{k=1}^\infty\phi(k) \Delta a  \\
&\approx& \frac{8}{3\sqrt{3}}\fracbrac{a'}{a'-a}^2\sqrt{\alpha\mu'}\sum_{k=1}^\infty \phi(k)|s_{k}(e/\eX)|^{1/2}\label{eq:tau_sum}
\end{eqnarray}
where $\phi(k)$ (called `Euler's totient function') gives the number of  $k$th order resonances contained in $\delta a$. The Euler totient function is defined as the number of integers up to $k$ that are relatively prime to $k$. 
To see that this is equivalent to the number of $k$th-order resonances within $\delta a$ consider the following:
the period ratios of all $k$th order resonances between the $J$:$J-1$ and $J+1$:$J$ first-order resonances (inclusive) can be written as $\frac{P}{P'}=\frac{(J-1)k+l}{Jk+l}$ with $0\le l\le k$. 
{Therefore, of the $k+1$ possible values for $l$, we should only retain  those that are relatively prime to $k$;  otherwise, the numerator and denominator are commensurate, and the  period ratio is
the same as one  of lower-order.}

\begin{figure}
    \centering
    \includegraphics[width=\columnwidth]{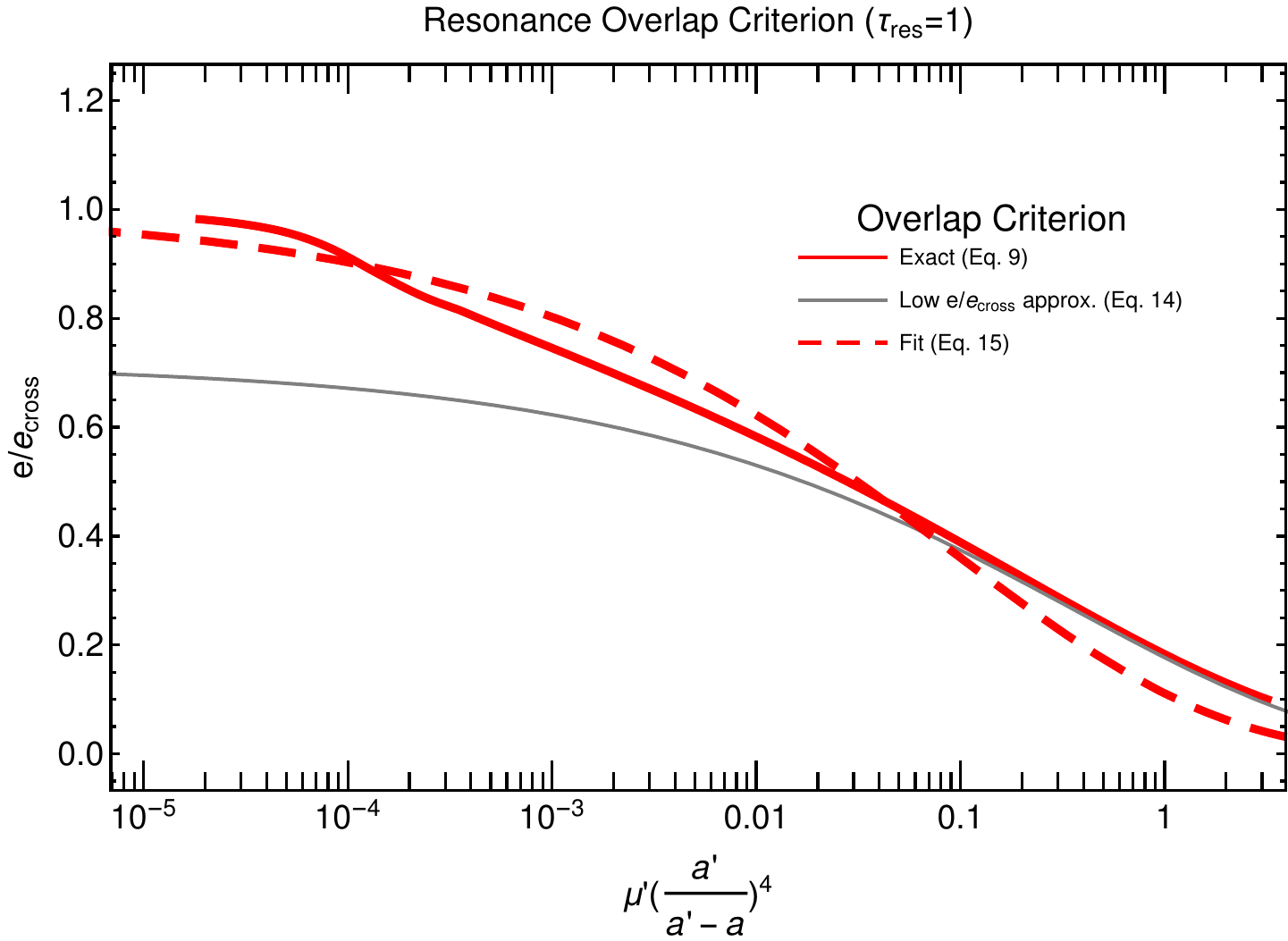}
    \caption{
    The critical eccentricity at which $\tau_\text{res}=1$ and the onset of chaos occurs as a function of $\mu'\fracbrac{a'}{a'-a}^4$.
    The solid red line shows the critical eccentricity computed by numerically solving Equation \eqref{eq:tau_sum} for $\tau_\text{res}=1$. The gray line shows the approximation Equation \eqref{eq:ecrit_approx_1} valid for $\mu'\fracbrac{a'}{a'-a}^4\gtrsim0.03$.
    The dashed red line shows our numerically fit approximation, Equation \eqref{eq:ecrit_approx_2}.
    }
    \label{fig:tau_compare}
\end{figure}

Our resonance overlap criterion, $\tau_{\rm res}=1$, provides  the critical $e/\eX$ for the onset of chaos
as a function of $\mu'$ and $a'/a$.
Figure \ref{fig:tau_compare} (thick red line) plots the critical eccentricity at which $\tau_{\rm res}=1$. \footnote{To evaluate the sum in Equation (\ref{eq:tau_sum})  we truncate at a finite value $k_{\rm max}$ such that, for each $e/\eX$, the sum increases by no more than 1\% upon doubling the number of terms.  We find that $k_{\rm max}\leq 1024$ is sufficient for eccentricities $e<0.99\eX$. }

Both spacing and mass determine the critical eccentricity, but only in the combination $\fracbrac{a'-a}{a'}/\mu'^{1/4}$;
in other words, the relevant spacing is in units of 
$\mu'^{1/4}a'$ rather than, e.g.,  number of Hill radii ($\mu'^{1/3}a'$). 
As is physically plausible, when the planet's mass is very small ($\mu'\rightarrow 0$), the test particle's $e$ must be
 very close to $e_{\rm cross}$ before
chaos is triggered---regardless of spacing.

\begin{figure}
    \centering
    \includegraphics[width=\columnwidth]{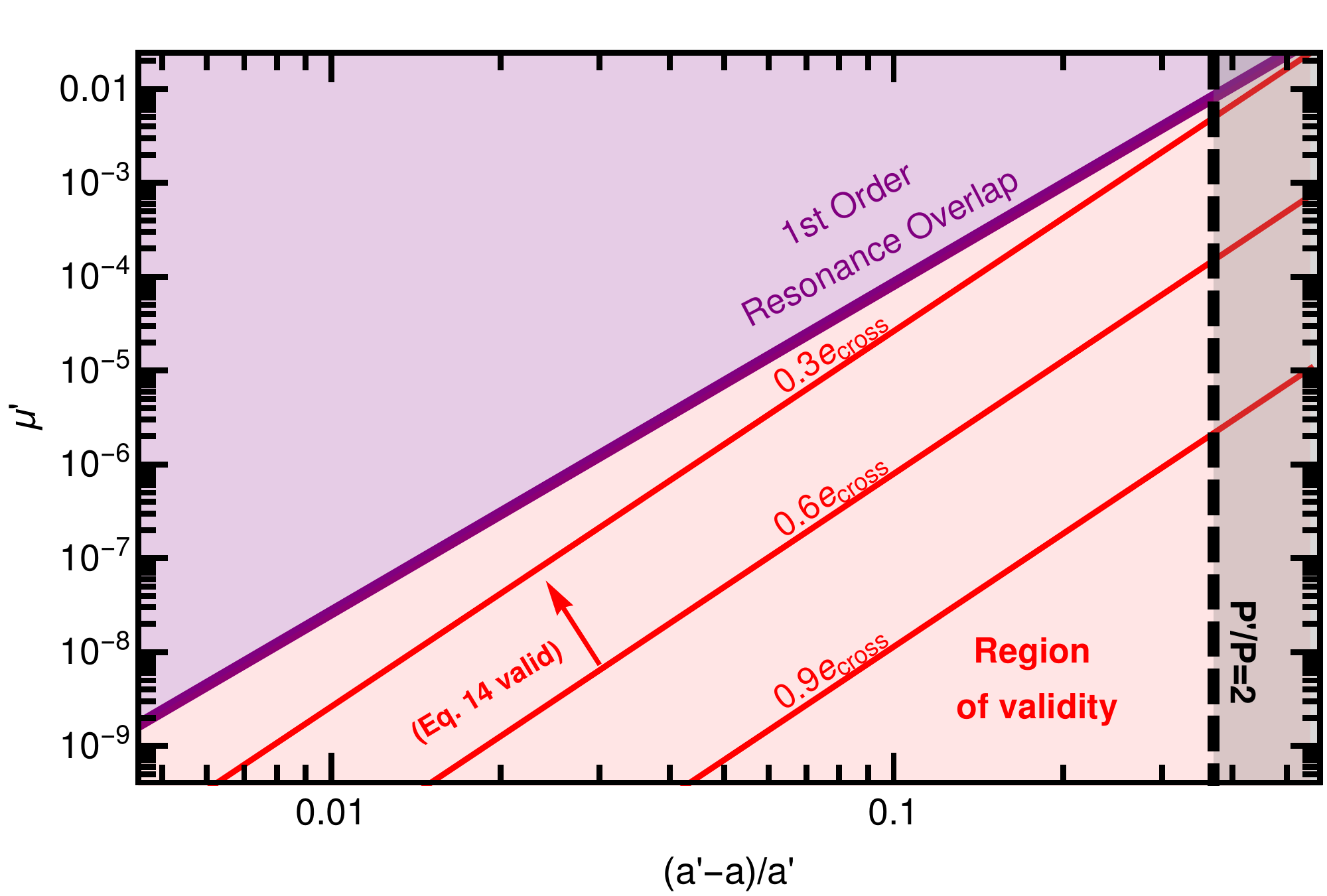}
    \caption{
    \label{fig:validity_regions}
    Onset of resonance overlap, and hence chaos, as a function of perturber mass and planet spacing. 
    The red lines show our overlap criterion from setting $\tau_{\rm res}=1$ in Equation \eqref{eq:tau_sum} at three values of critical eccentricity, as labeled.
    The purple line shows the onset of chaos due to the overlap of first-order mean motion resonances at small eccentricity according to \citet{Wisdom80}'s $\mu^{2/7}$ criterion. 
    Our overlap criterion, which predicts the critical eccentricity for the onset of chaos, applies below this line.
   The black dashed line indicates the 2:1 resonance.  Since our result for the critical eccentricity shown in Fig. \ref{fig:tau_compare} adopts the ``close approximation'' (see subsection \ref{sec:close_approx}), it becomes quantitatively incorrect
   beyond the 2:1. 
    }
\end{figure}

Figure \ref{fig:validity_regions}  (red lines) shows the spacing and mass dependencies separately.
From this figure, we see, for example, that planetary systems that have $(a'-a)/a'\sim 0.1$ and $\mu'\sim 10^{-5}$ (typical values for systems discovered 
by the {\it Kepler} telescope) have a
critical $e$ for chaos of $\sim 0.35\eX\sim 0.035$.  (Making both bodies massive---rather than working in the test particle limit---changes this number by of order unity; see Sec. \ref{sec:two_planet_overlap:generalize}). 

As mentioned above, our overlap criterion ignores the finite width of first-order resonances at small eccentricity.
\citet{Wisdom80} shows that these resonances overlap when 
  $\frac{a'-a}{a'}<1.46\mu'^{2/7}$.
  We plot this critical spacing as the purple line in Figure \ref{fig:validity_regions}.
 Our criterion for the critical eccentricity is only valid below this line, i.e.,  for masses $\mu'\le 0.27\fracbrac{a'-a}{a'}^{7/2}$.
 For larger masses,  chaos from first-order resonance overlap is expected at all eccentricities.

\subsection{Analytical Expressions for the Critical $e$}
\label{sec:two_planet_overlap:analytic_expression}
While it is straightforward to numerically evaluate Equation \eqref{eq:tau_sum}, it is useful to have an explicit formula 
for the critical $e$. 
To that end, we expand $s_k$ at low $e$, in which limit
\begin{equation}
    |s_{k}(e/\eX)| \approx \frac{\sqrt{3}\exp(k/3)}{\pi k}\fracbrac{e}{\eX}^k \label{eq:sk_leading_order}
\end{equation}
(see Eq. \eqref{eq:wk_approx2} 
 in Appendix \ref{sec:appendix}) after neglecting higher-order terms in $e/\eX$. Using the approximation $\phi(k)\approx \frac{6k}{\pi^2}$ (valid at large $k$),
 the sum in Equation \eqref{eq:tau_sum} becomes
 \begin{eqnarray}
 \sum_{k=1}^\infty \phi(k)|s_{k}(e/\eX)|^{1/2} &\approx& { 3^{1/4} 6\over\pi^{5/2}} \sum_{k=1}^\infty \sqrt{k}e^{k/6}\left( {e\over \eX} \right)^{k/2} \label{eq:ksum} \\
 &\approx& \frac{3^{1/4}6}{\pi^{5/2}}\int_{0}^{\infty} \sqrt{k}x^{k}dk\nonumber \\
&=&\frac{3^{5/4}}{\pi ^2 |\log \left(x\right)|^{3/2}}~.\label{eq:tau_sum_approx}
 \end{eqnarray}
after 
  defining $x\equiv\sqrt{\frac{e}{\eX}}\exp(1/6)$, and replacing the sum with an integral.\footnote{
  {Note that the $k$'s that dominate the integral leading to  Equation (\ref{eq:tau_sum_approx})  are  $k\sim {1\over \ln (\eX/e)}$. Hence, low-order resonances are the most important ones when $e\ll \eX$, and higher orders become the dominant ones at higher $e$.}}
    Inserting this
   into Equation \eqref{eq:tau_sum} and solving for the critical value of $e$ that gives $\tau_\text{res}=1$, we find
\begin{equation}
    e\approx 0.72\eX \exp\left[-1.4\mu'^{1/3}\fracbrac{a'}{a'-a}^{4/3}\right]~.\label{eq:ecrit_approx_1}
\end{equation}
Equation \eqref{eq:ecrit_approx_1} is compared with the numerically computed critical eccentricity in Figure  \ref{fig:tau_compare}.  We see that they agree well at $e/\eX\lesssim 0.6$, or equivalently for $\mu'\fracbrac{a'}{a'-a}^4\gtrsim0.03$. 

However, beyond this limit  the agreement is poor.   For example, in the limit $\mu'\rightarrow 0$  Equation \eqref{eq:ecrit_approx_1} mistakenly predicts that the onset of chaos occurs at $e=0.72\eX$  rather than the expected limit, $e=\eX$.
The error arises because Equation \eqref{eq:sk_leading_order} over-predicts  $|s_k|$, and hence resonance widths, for large $k$ when $e\gtrsim0.6\eX$.
Nonetheless, we obtain an adequate fit to numerical results over the full range of $0<e<\eX$ by adopting the functional form of Equation \eqref{eq:ecrit_approx_1} but dropping the factor of $0.72$ so that the appropriate $\mu'\rightarrow 0$ limit is recovered. Fitting for a new numerical constant in the exponential, we find that the formula
\begin{equation}
    e\approx \eX \exp\left[-2.2\mu'^{1/3}\fracbrac{a'}{a'-a}^{4/3}\right]  , \label{eq:ecrit_approx_2}    
\end{equation}
 plotted in Figure \ref{fig:tau_compare}, provides an acceptable approximation for the critical eccentricity yielding relative errors $<10\%$  when $\fracbrac{a'}{a'-a}^{4}\mu'<0.1$.

\subsection{Generalization to two massive  planets}
\label{sec:two_planet_overlap:generalize}
We  generalize our result for the threshold of chaos to the case of two massive planets, 
each of which may be eccentric. 
As will be shown in detail in \citet{Hadden_inprep}, the resonant dynamics of a massive planet pair can be cast in terms of a pendulum model almost identical to the one used in Section \ref{sec:two_planet_overlap:widths}.
The key step is the surprising fact that, to an excellent approximation, the resonant dynamics only depend on a single linear combination of the planet pairs' complex eccentricities, $e_1e^{i\varpi_1}$ and $e_2e^{i\varpi_2}$, where $e_i$ and $\varpi_i$ are the eccentricity and longitude of perihelion of the inner ($i=1$) and outer ($i=2$) planet.
{This represents a nontrivial generalization of a previously-derived result for first-order resonances \citep{Sessin1984,wisdom1986,Batygin2013,Deck2013overlap}.}
To oversimplify the results of Hadden slightly for the sake of clarity, the resonance dynamics depends 
only on the difference in complex eccentricities:
\begin{equation}
{\cal Z}={1\over\sqrt{2}}\left( e_2e^{i\varpi_2} -  e_1e^{i\varpi_1}\right)~. \label{eq:zdiff}
\end{equation}
We will refer to ${\cal Z}$ as the complex relative 
eccentricity and its magnitude as the relative eccentricity, which we will write as $Z\equiv|{\cal Z}|$.\footnote{\label{ft:zdef}
A more precise statement of Hadden's result is as follows:
 if we define the complex quantities
\begin{eqnarray}
\begin{pmatrix}{\cal Z} \\ {\cal W} \end{pmatrix} = \begin{pmatrix} \cos\theta &~-\sin\theta \\  \sin\theta &~\cos\theta \end{pmatrix} \begin{pmatrix} e_2e^{i\varpi_2} \\  e_1e^{i\varpi_1} \end{pmatrix}
\label{eq:zwdef}
\end{eqnarray}
where $\theta = \arctan[(a_1/a_2)^{0.37}]$ then the dynamics of nearby resonances will depend 
almost entirely on $\cal Z$ and are essentially independent of ${\cal W}$.  Throughout this paper $\cal Z$ really refers to that in the above matrix equation {rather than the oversimplified form of Equation \ref{eq:zdiff}};
but note that
for period ratios interior to the 2:1 resonance, $\theta$ differs from $\pi/4$ by no more than $10\%$ so that ${\cal Z}\approx \frac{1}{\sqrt{2}}\left(e_2e^{i\varpi_2} -  e_1e^{i\varpi_1}\right)$ provides an adequate approximation for most purposes. We will refer to  ${\cal W}\approx \frac{1}{\sqrt{2}}\left(e_2e^{i\varpi_2} +  e_1e^{i\varpi_1}\right)$ as the average complex eccentricity.
}

Resonant widths scale with mass and eccentricity in essentially the same way as in the test-particle case, Equation  \eqref{eq:da_ctp}, after replacing $e \rightarrow \sqrt{2}Z$ and $\mu'\rightarrow\mu_1+\mu_2$. 
Proceeding through exactly the same resonance optical depth formulation presented in Section \ref{sec:two_planet_overlap:optical_depth}
yields
\begin{multline}
\tau_\text{res}\approx \frac{8}{3\sqrt{3}}\fracbrac{a_2}{a_2-a_1}^2\sqrt{\alpha(\mu_1+\mu_2)}\sum_{k=1}^\infty \phi(k)\left|s_{k}\fracbrac{\sqrt{2}Z}{\eX}\right|^{1/2}
\label{eq:zcrit_exact}
\end{multline}
as the generalization of Equation \eqref{eq:tau_sum} when both planets are massive and/or eccentric. 
Similarly,
\begin{equation}
Z \approx \frac{e_\text{cross}}{\sqrt{2}}\exp\left[-2.2(\mu_1+\mu_2)^{1/3}\left(\frac{a_2}{a_2-a_1}\right)^{4/3}\right]
\label{eq:zcrit_approx}
\end{equation}
provides an approximate formula for the critical $Z$ for the onset of chaos as the generalization of Equation \eqref{eq:ecrit_approx_2}.

{When both planets are massive and/or eccentric, $Z$ is not a  strictly conserved quantity, but rather can vary on secular timescales. In principle, this means that planet pairs can evolve secularly from regions of phase space where resonances are initially not overlapped into overlapped regions. In practice, however, secular variations in $Z$ are generally negligible because the linear combination of complex eccentricities that defines $Z$ (Equation \ref{eq:zdiff}) is nearly identical to one of the secular eigen-modes of the two-planet system. The secular evolution of $Z$ will be explored further by \citet{Hadden_inprep}. }
\section{Comparison with Numerical Results}
\label{sec:numerical_compare}
\begin{figure}
\includegraphics[width=1\columnwidth]{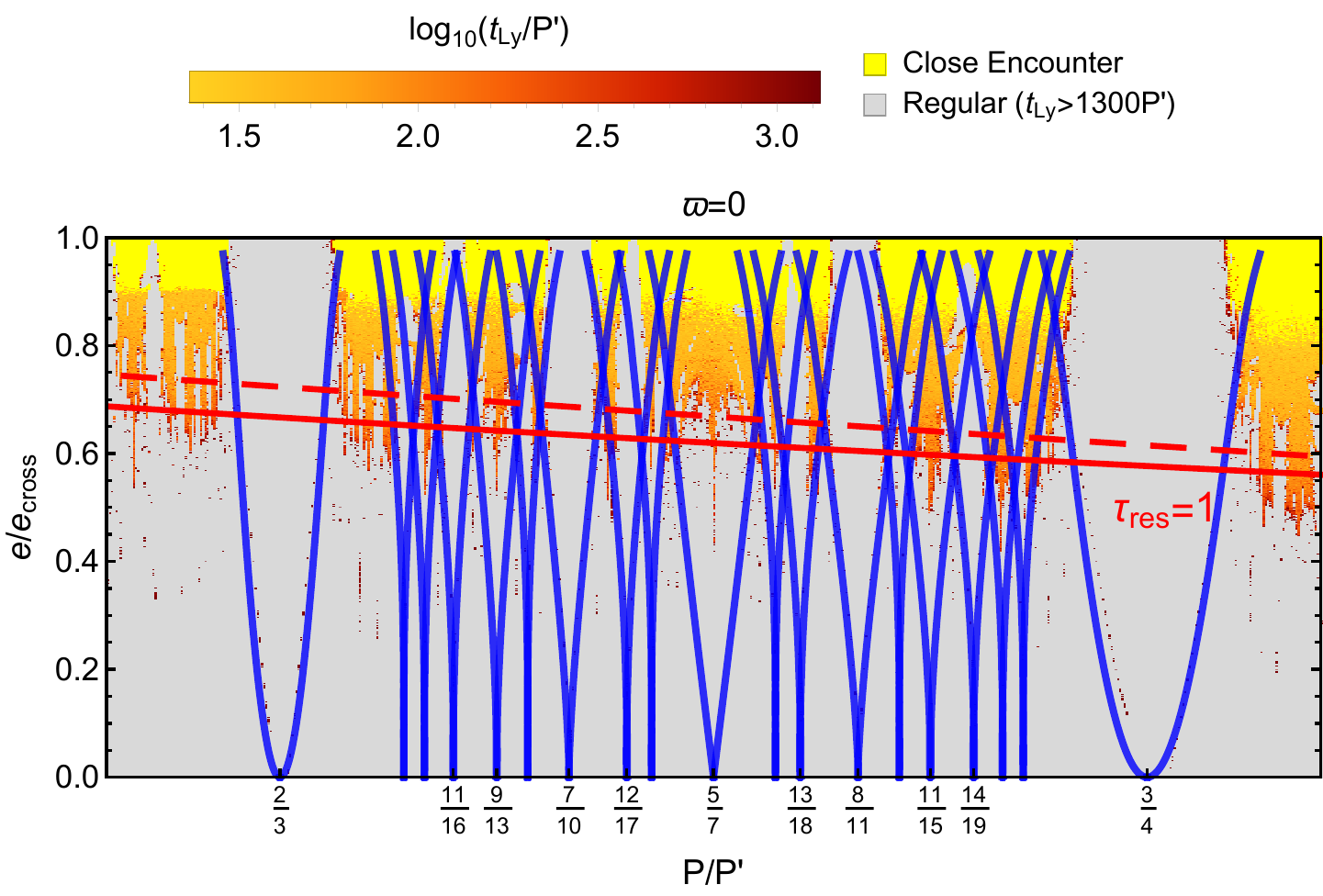}
\includegraphics[width=1\columnwidth]{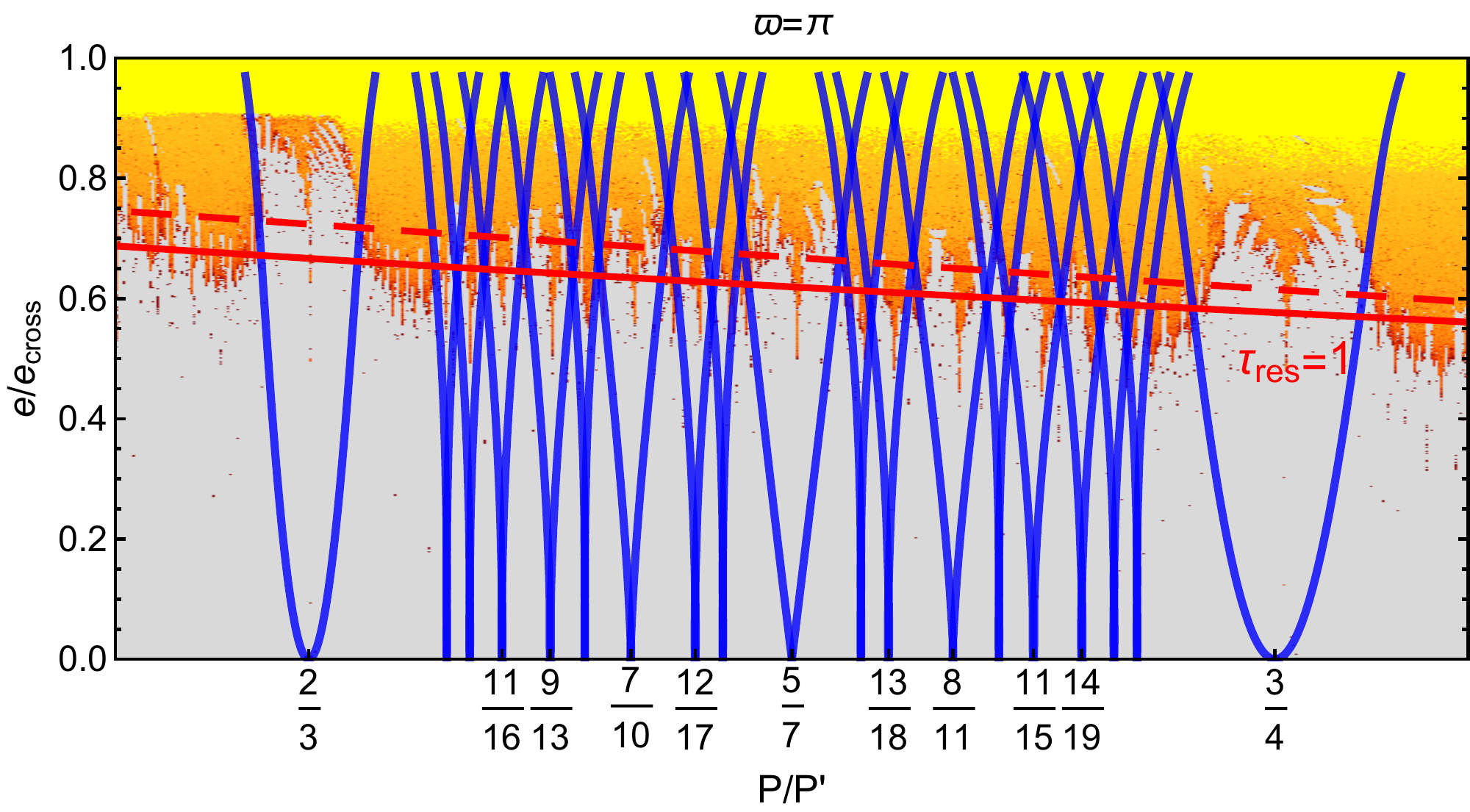}
\centering
\caption{\label{fig:chaos_compare}
Chaotic structure of phase space as a function of period ratio and  $e/\eX$  for a test-particle subject to an exterior  circular perturber with mass $\mu'=10^{-5}$.  
The color scale indicates the Lyapunov time of chaotic trajectories computed for a grid numerical integrations (see text for initial conditions). 
Integrations that led to a close encounter (within one Hill radius) were stopped early and are marked in bright yellow.
Initial conditions determined to have $t_\text{Ly}>1300P'$ are assumed regular and marked as gray.
The blue separatrices and our red overlap criteria are repeated from Figure \ref{fig:optical_depth_schematic}. 
}
\end{figure}
We compare the prediction of our resonance overlap criterion with the results of numerical integrations in Figure \ref{fig:chaos_compare}.
All numerical integrations are done with the WHFast integrator \citep{RTwhfast2015} { based on the symplectic mapping algorithm of \citet{wisdom_holman1991} and} implemented in the REBOUND code \citet{RL12}. {Integration step sizes are set to 1/30th of the orbital period of the inner planet unless stated otherwise. To ensure that our results are not driven by numerical artifacts of our integration method, in Appendix \ref{sec:appendix:integrator_compare} we compare results derived using the WHFast integrator with results obtained using the high-order, adaptive time step  IAS15 routine \citep{RS15} implemented in the REBOUND code. The two methods show excellent agreement, indicating that our results are not affected by numerical artifacts.}

Initial conditions for {the top panel} are chosen {to be}  $\lambda=\lambda'=0$ and {$\varpi=0$}; {the bottom panel is the same, but with $\varpi=\pi$}. The integrations lasted for 3000 planet orbits. 
To compute  Lyapunov times we used 
the MEGNO chaos indicator \citep{2003PhyD..182..151C}  built into REBOUND.
The MEGNO grows linearly at a rate of $1/t_\text{Ly}$, where $t_{Ly}$ is the Lyapunov time, for chaotic trajectories while asymptotically approaching a value of 2 for regular trajectories.
Throughout the paper we report $t_\text{Ly}$ values estimated by simply dividing integration runtimes by MEGNO values.
In the figure, trajectories with $t_\text{Ly}>1300P'$  are considered regular and plotted in gray.
We are unable to detect chaos for initial conditions with longer Lyapunov times given the limited duration of our integrations. 
However, we find that longer integration runtimes, up to $10^6$ orbits, do not significantly change the number of simulations classified as chaotic.

Figure  \ref{fig:chaos_compare} shows that the analytic overlap criterion ($\tau_{\rm res}=1$) broadly agrees with the $N$-body results, predicting the transition to large-scale chaos as a function of  eccentricity in the period range shown.
{The boundaries between regular and chaotic orbits  in the top and bottom panels are similar, demonstrating that the onset of resonance overlap does not depend strongly on the initial orbital phase.}
There are at least two caveats to our overlap criterion: first,
non-chaotic regions extend above the predicted overlap region in the top panel, most prominently for the first-order 3:2 and 4:3 resonances, but also at other odd-ordered MMRs.  
Our choice of initial conditions in the top panel of Figure \ref{fig:chaos_compare} places the test particle near stable fixed points of these odd-order MMRs and  regular regions of phase-space clearly remain near these fixed points even when the resonances are overlapped. 
Second, the curve $\tau_{\rm res}=1$ is not a sharp boundary. A mixture of chaotic and regular trajectories is generically expected in regions of marginal resonance overlap and the boundary between regular and chaotic phase-space exhibits fractal structure \citep[e.g.,][]{LLBook}.
Nonetheless, the heuristic resonance overlap criterion provides an excellent prediction for onset of chaos from a coarse-grained perspective.

\begin{figure*}
\centering
\includegraphics[width=0.33\textwidth]{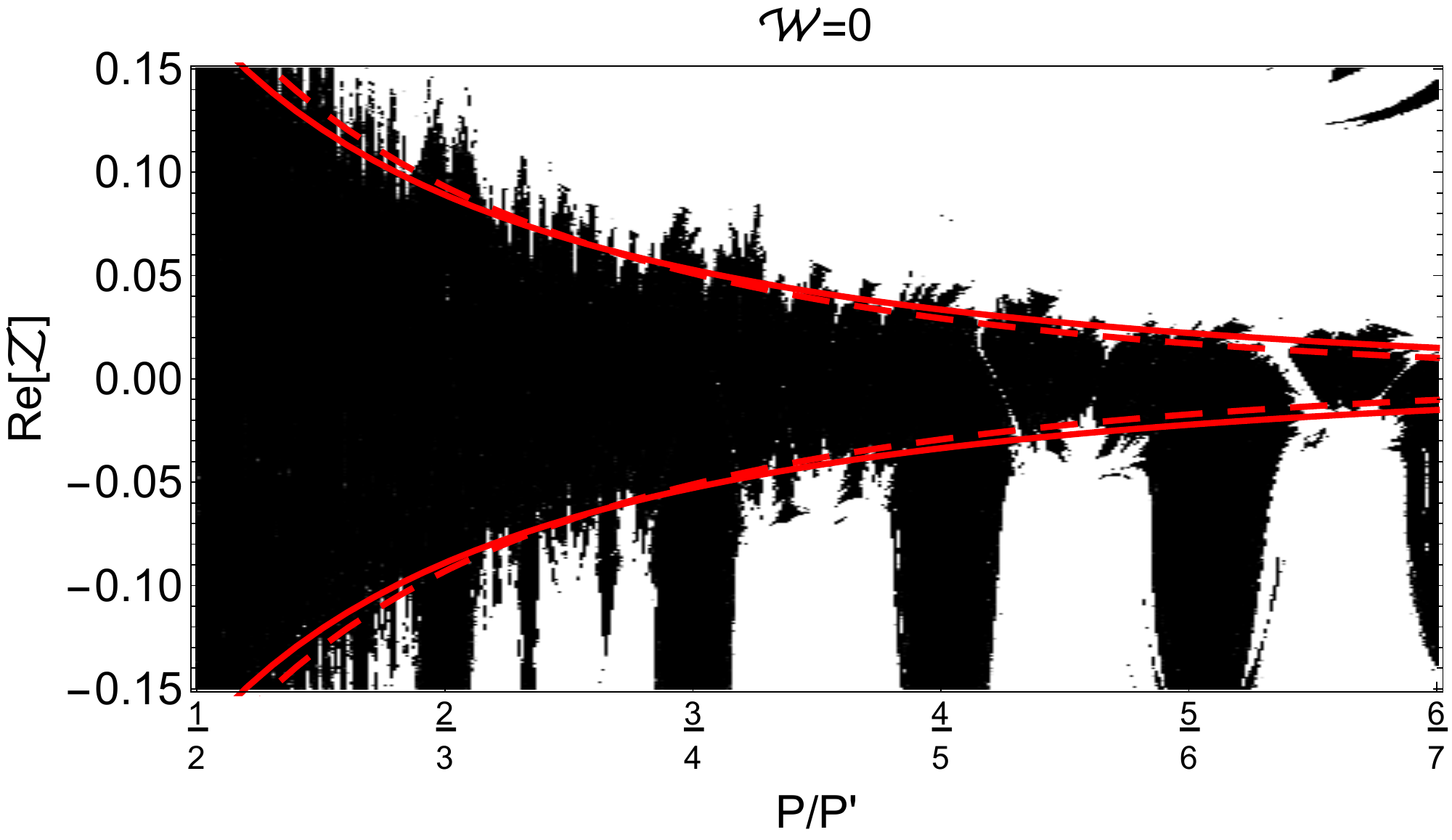}
\includegraphics[width=0.33\textwidth]{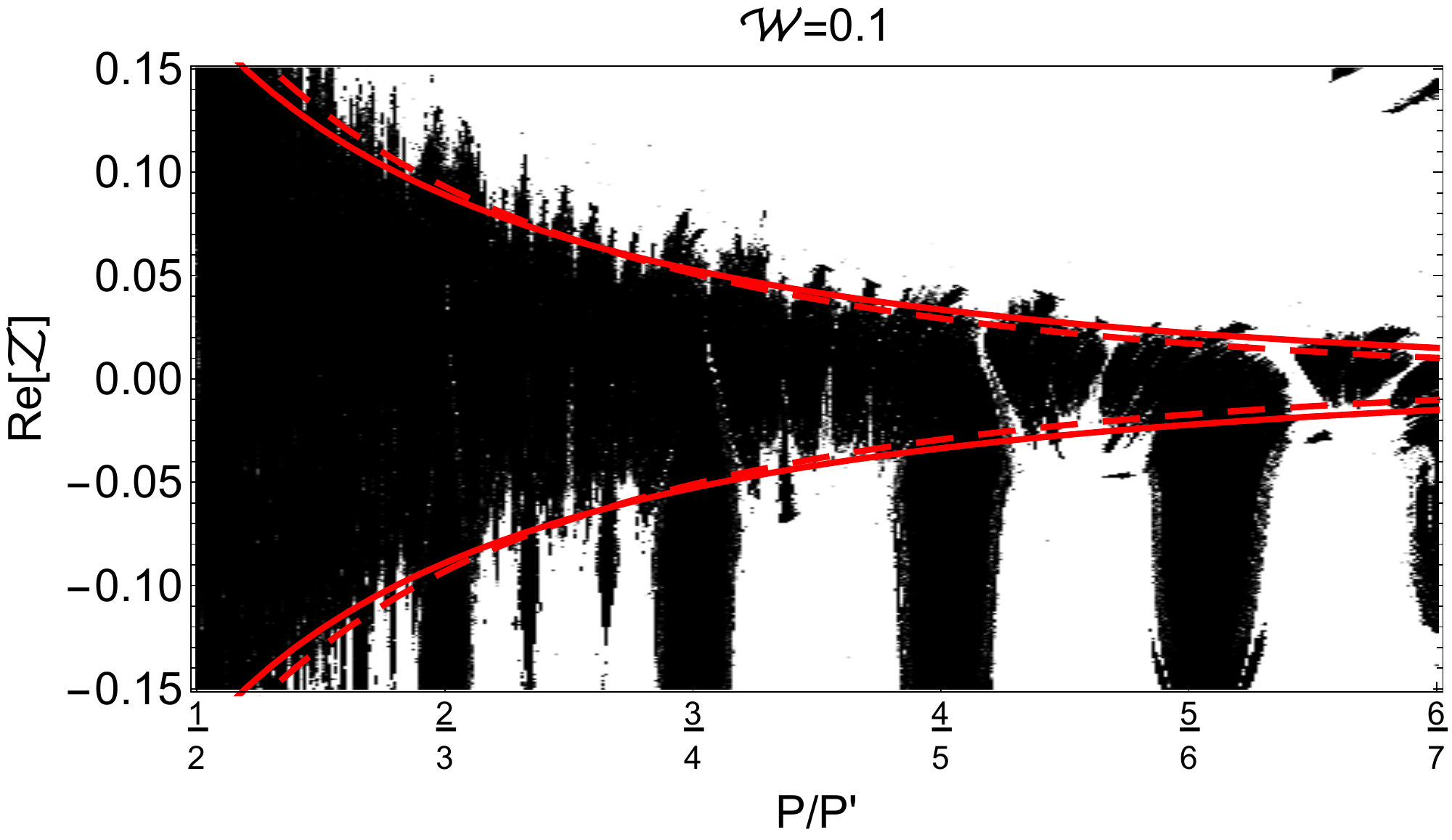}
\includegraphics[width=0.33\textwidth]{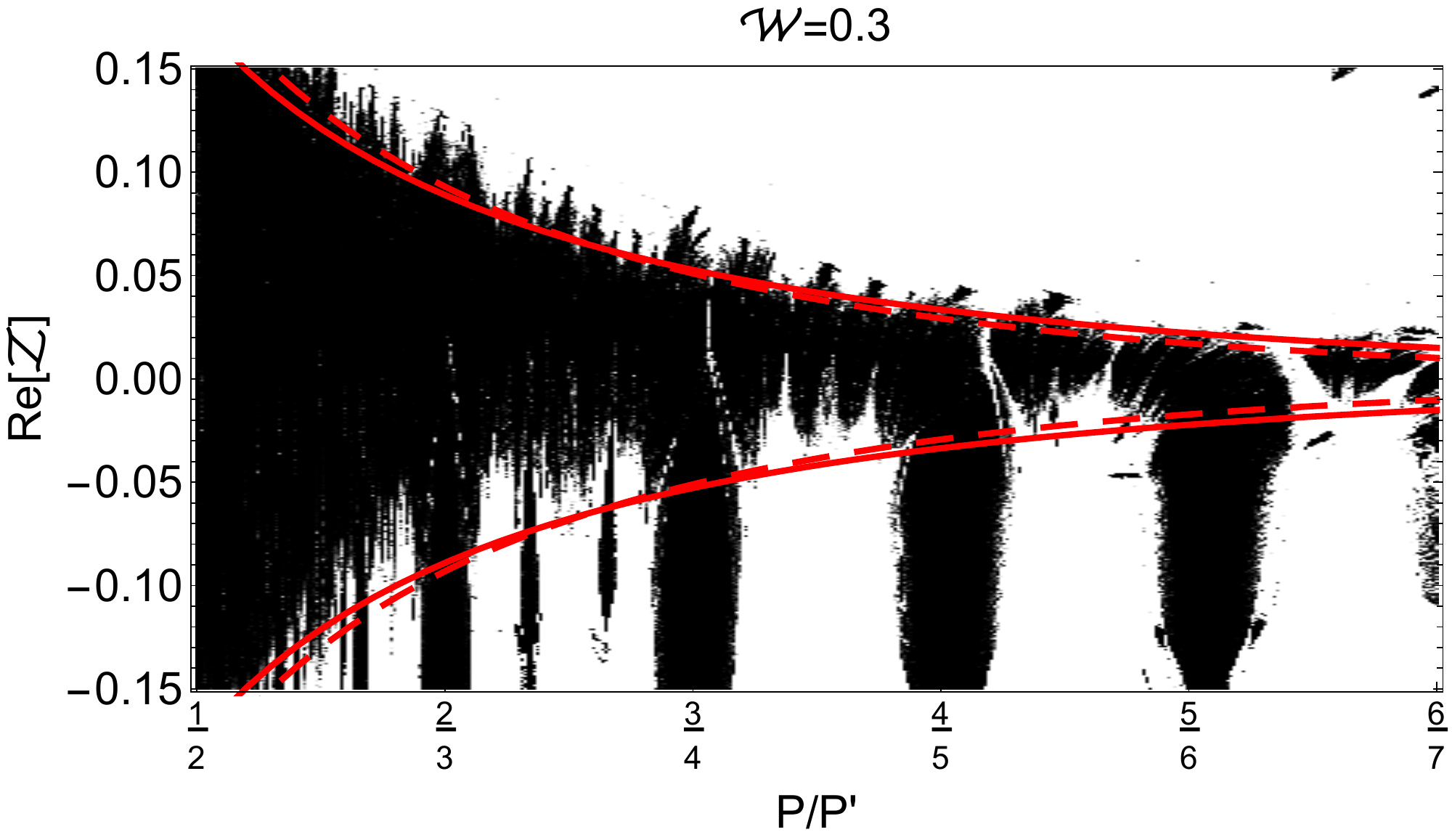}
\caption{
\label{fig:w_compare} Chaotic phase space structure for two planets of mass $m_1=m_2=3\times10^{-5}M_*$ with different values of ${\cal W}$ (Equation \ref{eq:zwdef}) integrated for 3000 orbits of the outer planet. 
{The planets are initialized with $\lambda_1=\lambda_2=0$ and $\text{Im}{\cal Z}=0$.}
Chaotic trajectories are shown in white while regular trajectories appear black.
{More precisely}, the background consists of a gray-scale spanning a narrow range of MEGNO values ranging from MEGNO$\le 2$ (black; $t_{Ly}\gtrsim 1500  P_2$) to MEGNO$\ge 6$ (white; $t_{Ly}\lesssim 500 P_2$).
The narrow gray-scale range emphasizes the sharp transition from regular to chaotic orbits.
Integrations that led to a close encounter within one Hill radius were stopped early and are also marked in white. 
The vertical axis shows the real component of the initial complex relative eccentricity ${\cal Z}$; the imaginary component is zero. 
The onset of chaos predicted by Equations \eqref{eq:zcrit_exact} and \eqref{eq:zcrit_approx} are shown as red solid and dashed lines, respectively.
The predicted onset of chaos does not depend on ${\cal W}$ and is the same in all three panels.
}
\end{figure*}

 Figure  \ref{fig:w_compare}  shows  results for two massive planets. 
As we have argued, the threshold for chaos should depend on planet eccentricities only through the {\it relative} complex eccentricity $\cal Z$, and not on the {\it average} complex eccentricity $\cal W$ (see footnote \ref{ft:zdef}) 
To test this, each panel of Figure \ref{fig:w_compare} displays numerical results on a grid computed from initial conditions that are identical except in their initial value of $\cal W$. In all three cases, the boundary of chaos agrees quite well with the theoretical prediction. Even 
when ${\cal W}=0.3$, which is significantly bigger than the relative eccentricity in the plot, there is only a modest effect on the stability boundary seen in the simulations. {Because the planets can have significant eccentricities when ${\cal W}$ is large, we use reduced time steps for the integrations shown in Figure \ref{fig:w_compare}. The step size is chosen based on the initial eccentricity of the inner planet to be $\frac{1}{30}\left({2\pi}/{\dot{f}_p}\right)$, where $\dot{f}_p$ is the time derivative of the planet's true anomaly at pericenter.  }

\begin{figure}
\includegraphics[width=\columnwidth]{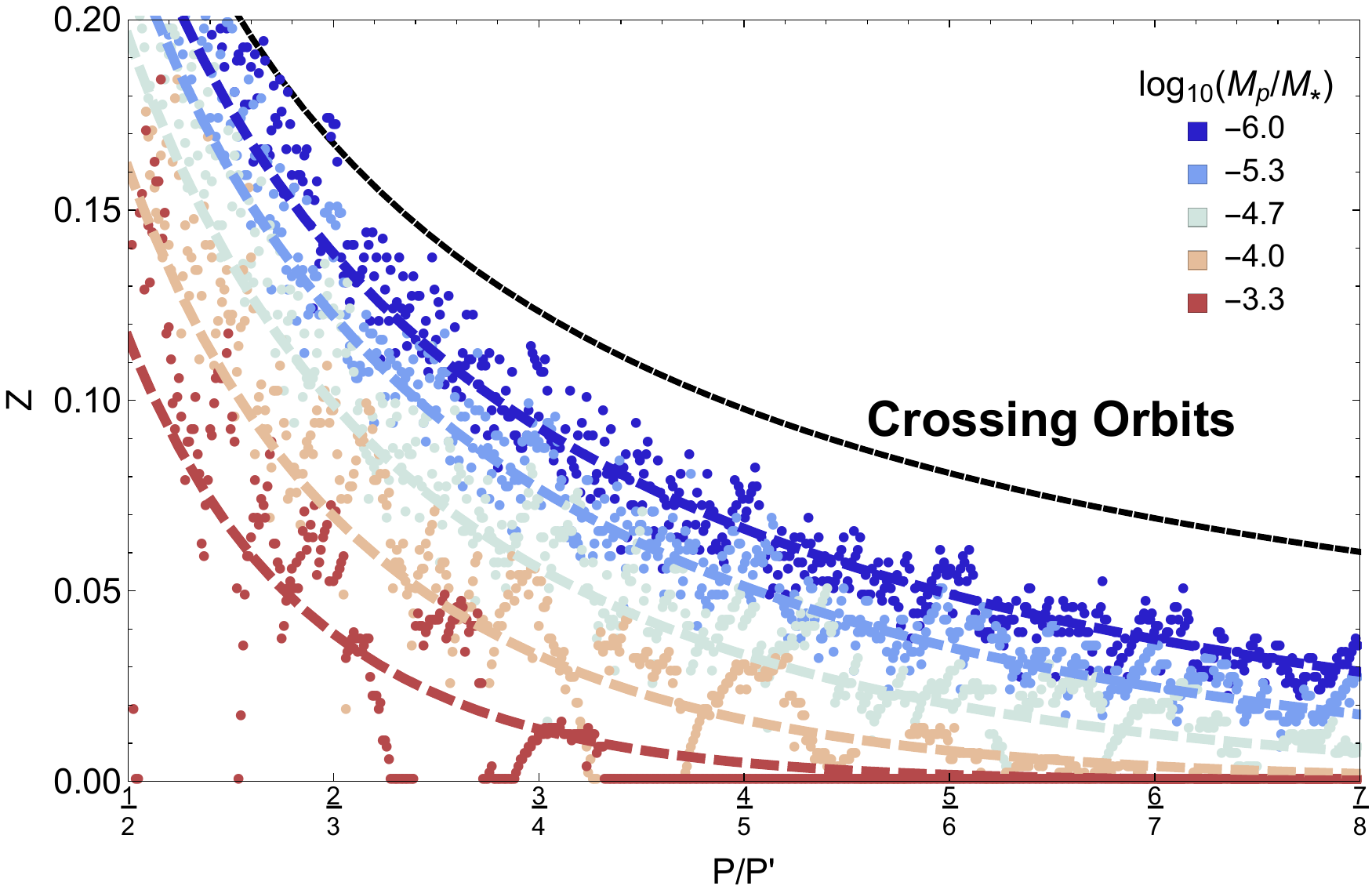}
\centering
\caption{\label{fig:optical_depth} Critical eccentricity as a function of planet period ratio for two equal mass planets using the $\tau_{res}=1$ criterion (see main text).
Points show the transition to chaos computed from $N$-body simulations.
Analytic predictions given by Equation \eqref{eq:zcrit_approx} are plotted as dashed lines. 
Curves  and points are colored according to planet mass. 
The value of $Z$ corresponding to crossing orbits is indicated by the black dashed line.
}
\end{figure}

\begin{figure}
\includegraphics[width=\columnwidth]{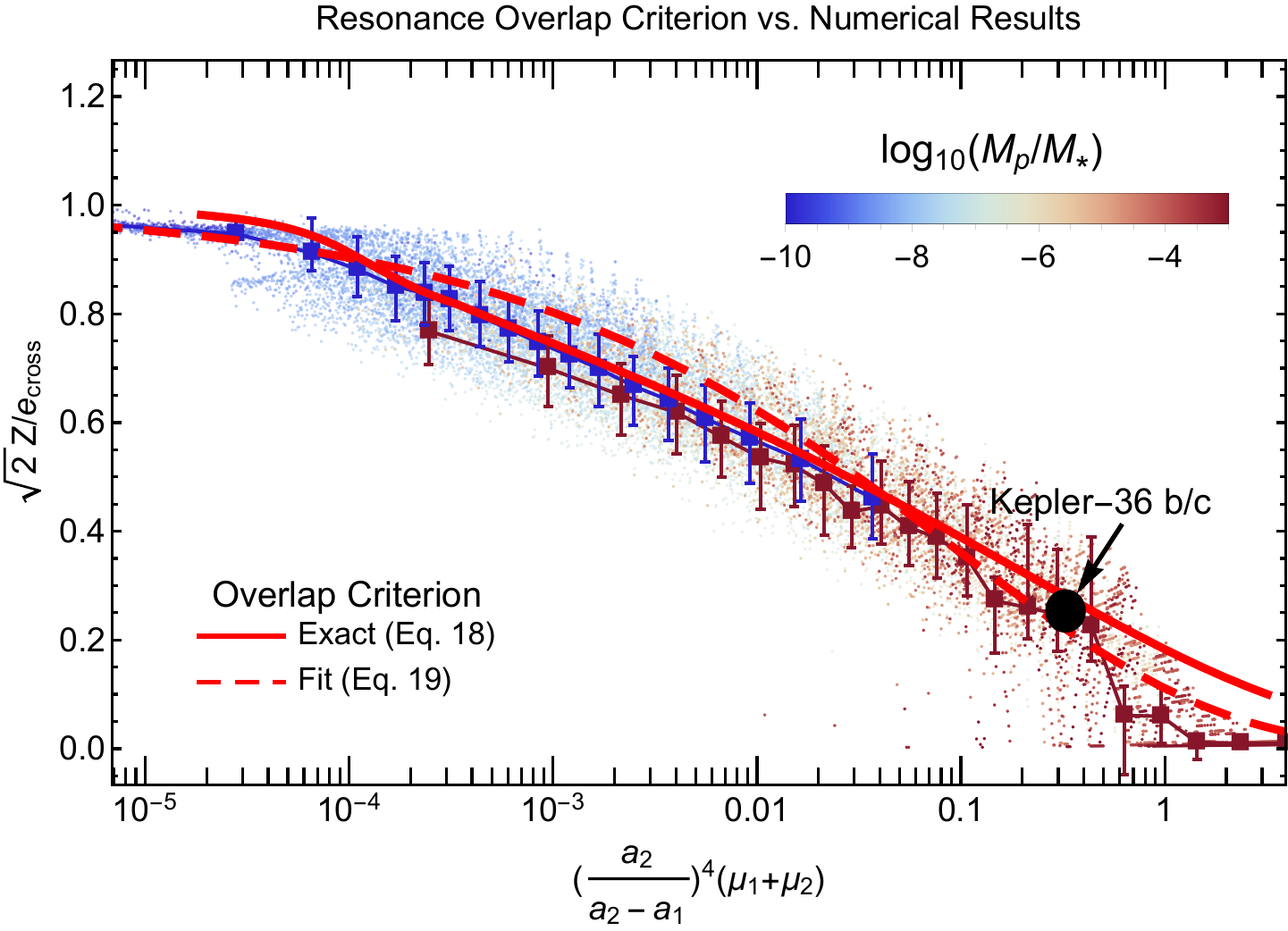}
\centering
\caption{\label{fig:optical_depth_2} 
Critical $Z$ as a fraction of $e_\text{cross}/\sqrt{2}$, plotted versus the quantity $(a/\Delta a)^4(\mu_1+\mu_2)$ computed from numerical simulations over wide range of planet spacings and masses in the range $[10^{-10}-10^{-3}]M_*$. 
Points are colored according to planet mass.  
The square markers and error bars indicate the median and $1\sigma$ range of binned numerical results for planet masses $>10^{-6}M_*$ (red) and $<10^{-6}M_*$ (blue).
Quantities for the planet pair Kepler-36 b and c are indicated by the black circle (see Section \ref{sec:chaos_vs_stability}).
}
\end{figure}

Figures \ref{fig:optical_depth} and \ref{fig:optical_depth_2} compare our overlap criterion with suites of numerical simulations with a wide range of planet masses and spacings.
The planets are equal mass and initial conditions are chosen so that $\lambda=\lambda'=0$, $\arg{\ZZ}=0$, and ${\cal W}=0$.
The transition to chaos is measured from numerical simulations by computing period-ratio/eccentricity grids similar to those shown in Figures \ref{fig:chaos_compare} and \ref{fig:w_compare} and identifying the minimum value of initial $Z$ for a given period ratio that yields chaos (taken to mean MEGNO$>5$ after a 3000 orbit integration, though our results are not sensitive to the choice of MEGNO threshold).
Figure \ref{fig:optical_depth} shows that the onset of chaos occurs at $Z$s that are a decreasing fraction of the orbit-crossing value as the planets' masses are increased.
In all cases, the numerical results broadly agree with our prediction.

Figure \ref{fig:optical_depth_2} confirms that the scaling of the critical eccentricity with planet mass predicted by the optical depth method holds over a wide range of planet masses and spacings. Note that in Figure \ref{fig:optical_depth_2} points computed from wide range of masses are plotted at every value of $(a/\Delta a)^4(\mu_1+\mu_2)$.
This figure also shows excellent overall agreement with the analytic predictions of Equations \eqref{eq:zcrit_exact} and \eqref{eq:zcrit_approx}.

\section{Onset of Chaos and Long-term Stability}
\label{sec:stability}
\subsection{Comparison with Other Stability Criteria}
\label{sec:criteria_compare}
\begin{figure*}
\centering
\includegraphics[width=.9\textwidth]{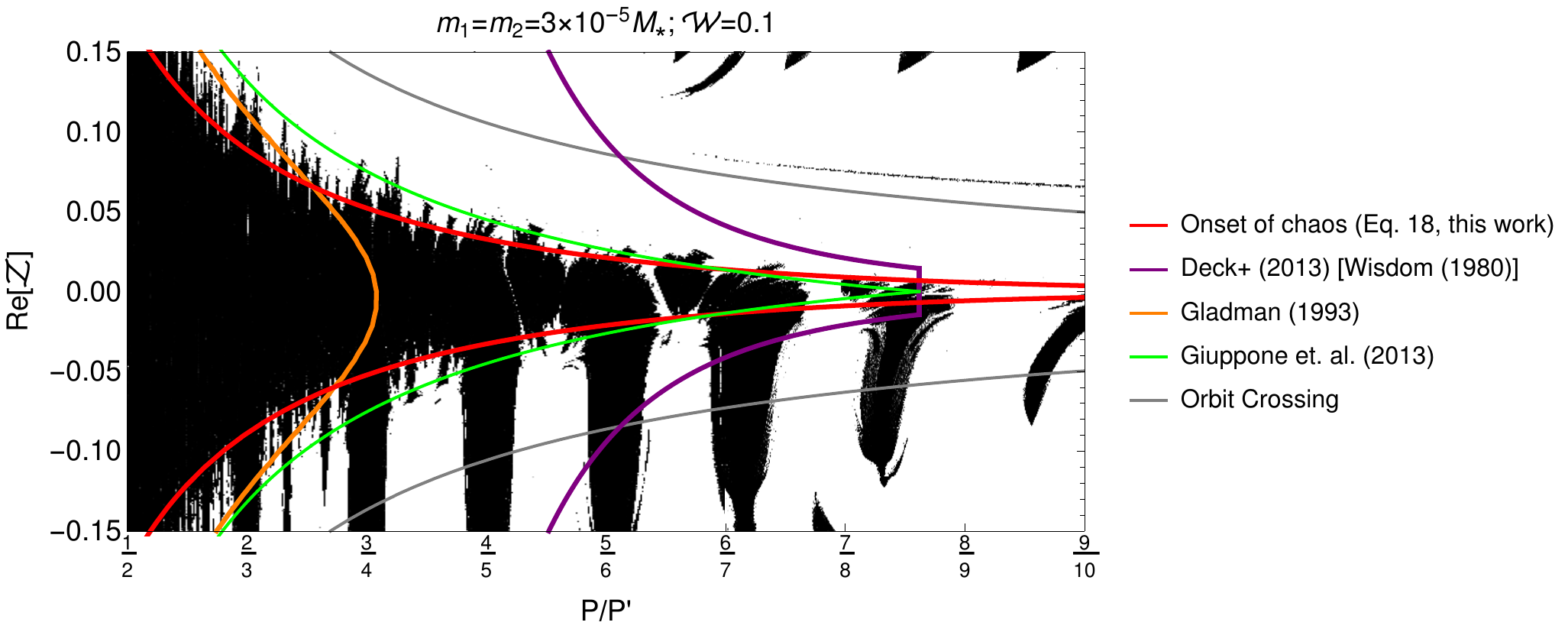}
\caption{\label{fig:criteria_compare} 
Chaotic phase space structure for a two-planet system with the same parameters as in the middle panel of Figure \ref{fig:w_compare}.
Different colored curves show the predictions of different stability criteria (see text). The gray curve shows  orbit-crossing eccentricity values. 
Note that the background grid of white and black points are plotted at their {\it initial} $P/P'$ and $e$. 
That is why the black resonant tongues toward the right of the figure do not line up with the nominal positions of the first-order MMR's. (They do line up when we plot mean, rather than initial, orbital elements.)
The resonance-overlap predictions have been corrected to account {for} the difference between the osculating and mean period ratio using a correction accounting for 0th order resonances. 
}
\end{figure*}

We compare our result for the onset of chaos in two-planet systems with some of the other stability criteria that appear
 in the literature (Figure \ref{fig:criteria_compare}).
  \citet{Wisdom80} derived a criterion for the onset of chaos based on resonance overlap for a   test-particle subjected to a planetary perturber, both of which are {nearly} circular. When $e$ {is sufficiently small}, only first-order resonances have non-vanishing width (see Section \ref{sec:two_planet_overlap:widths}). Therefore Wisdom considered only the overlap of first-order MMR's to derive his well-known $\mu^{2/7}$  criterion. \citet{Deck2013overlap} extended Wisdom's result to the case of two massive  (but still circular) planets, predicting the transition to chaos occurs at a critical spacing $(a_2-a_1)/a_1\approx1.46(\mu_1+\mu_2)^{2/7}$.  The vertical purple line at $P/P'$ slightly greater than 7/8 in Figure \ref{fig:criteria_compare}  shows \citet{Deck2013overlap}'s $\mu^{2/7}$ prediction. As seen in that figure,  their criterion works well at $Z=0$.  By contrast, since we ignore the peculiar low-$e$ behavior of first-order MMR's  our formula does not recover this result.    Therefore our threshold for chaotic onset  (Equation \eqref{eq:zcrit_approx}) should  be restricted to separations $(a_2-a_1)/a_1 \gtrsim (\mu_1+\mu_2)^{2/7}$. \citet{Deck2013overlap} also  account for  eccentricities in their overlap criterion by {generalizing the results of \citet{MustillWyatt2012} to include} the eccentricity-dependence of first-order MMR widths. 
  {\citet{MustillWyatt2012}'s criterion for the critical eccentricity can be stated as $e\propto \fracbrac{a-a'}{a'}^{4}\mu^{-1}\ecross$ \citep[see also ][]{cutler:2005}. Their result, based on the overlap of first-order resonances, can be recovered by considering only the $k=1$ term in the sum in Equation \eqref{eq:tau_sum} and noting that $s_1\propto e/\ecross$ when eccentricity is small. }
  Their prediction, {as generalized by  \citet{Deck2013overlap}}, is plotted as the non-vertical part of the purple curve in Figure \ref{fig:criteria_compare}; however, it significantly over-predicts the critical $e$ because it ignores MMR's with $k> 1$. 
 {In particular, the $k>1$ terms in Equation \eqref{eq:tau_sum} defining $\tau_\text{res}$ represent a fractional correction to the leading $k=1$ term of $>50\%$ for $e>0.09\eX$.}
  
\begin{figure*}
\centering
\includegraphics[width=0.3\textwidth]{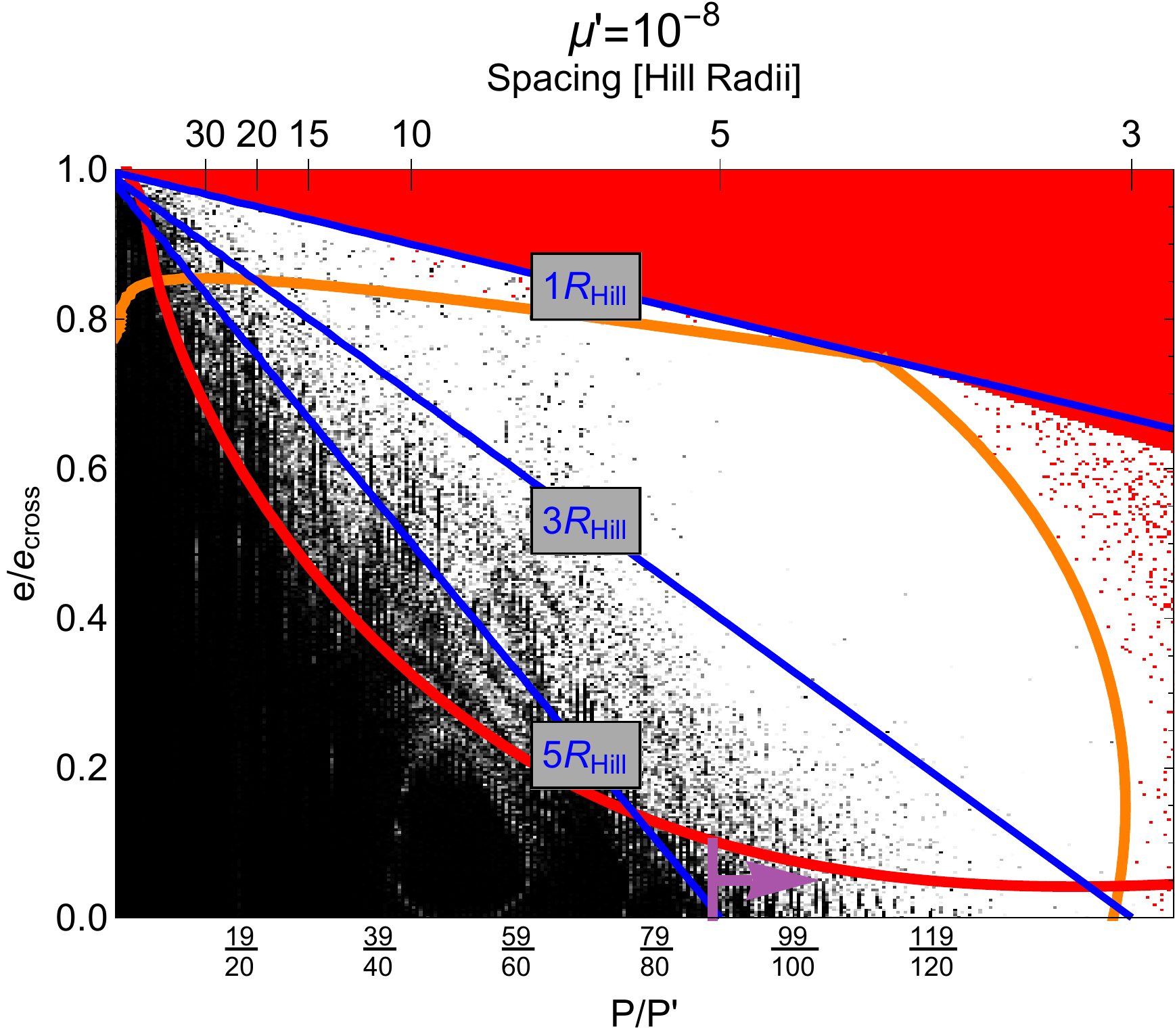}
\includegraphics[width=0.3\textwidth]{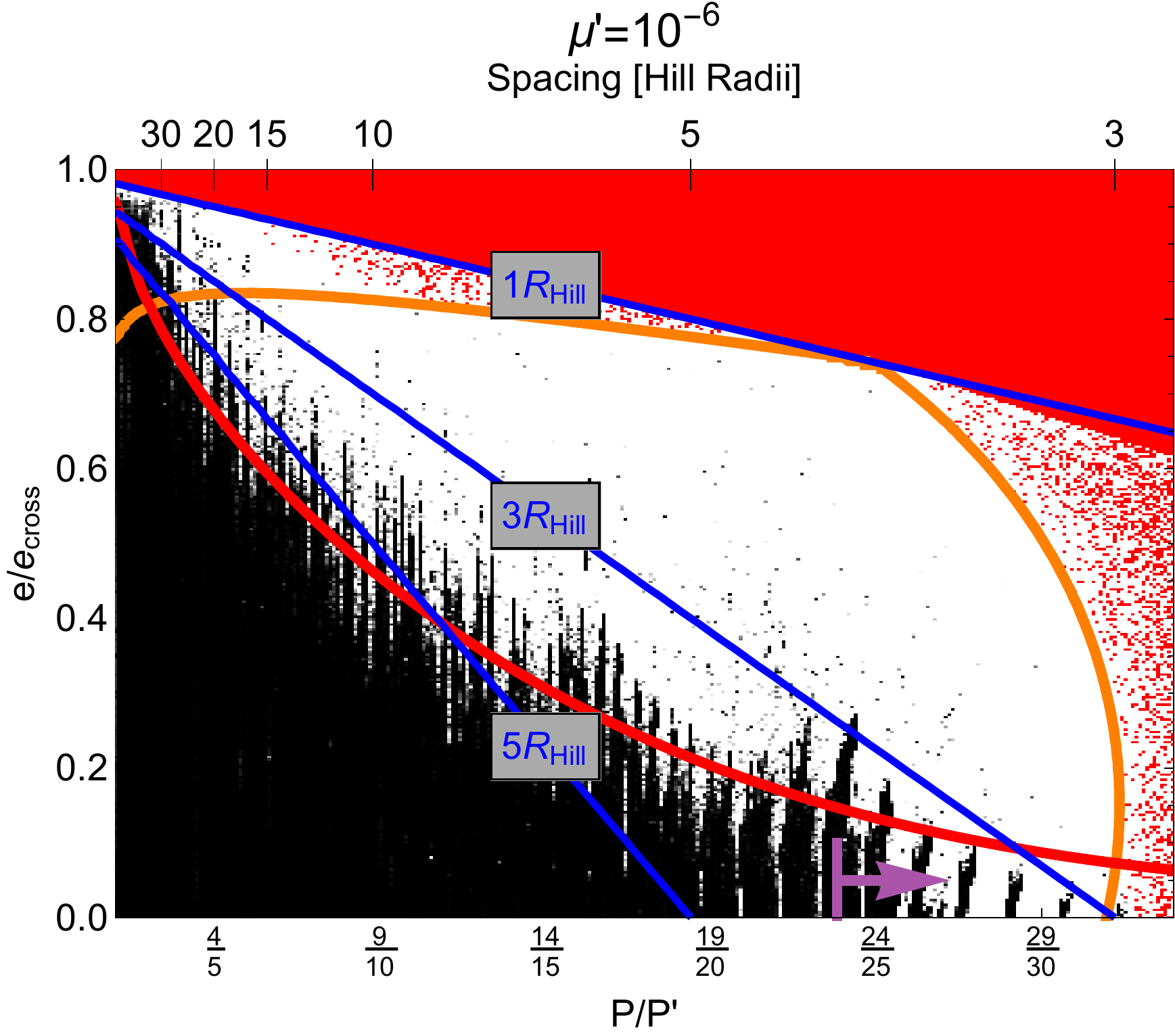}
\includegraphics[width=0.3\textwidth]{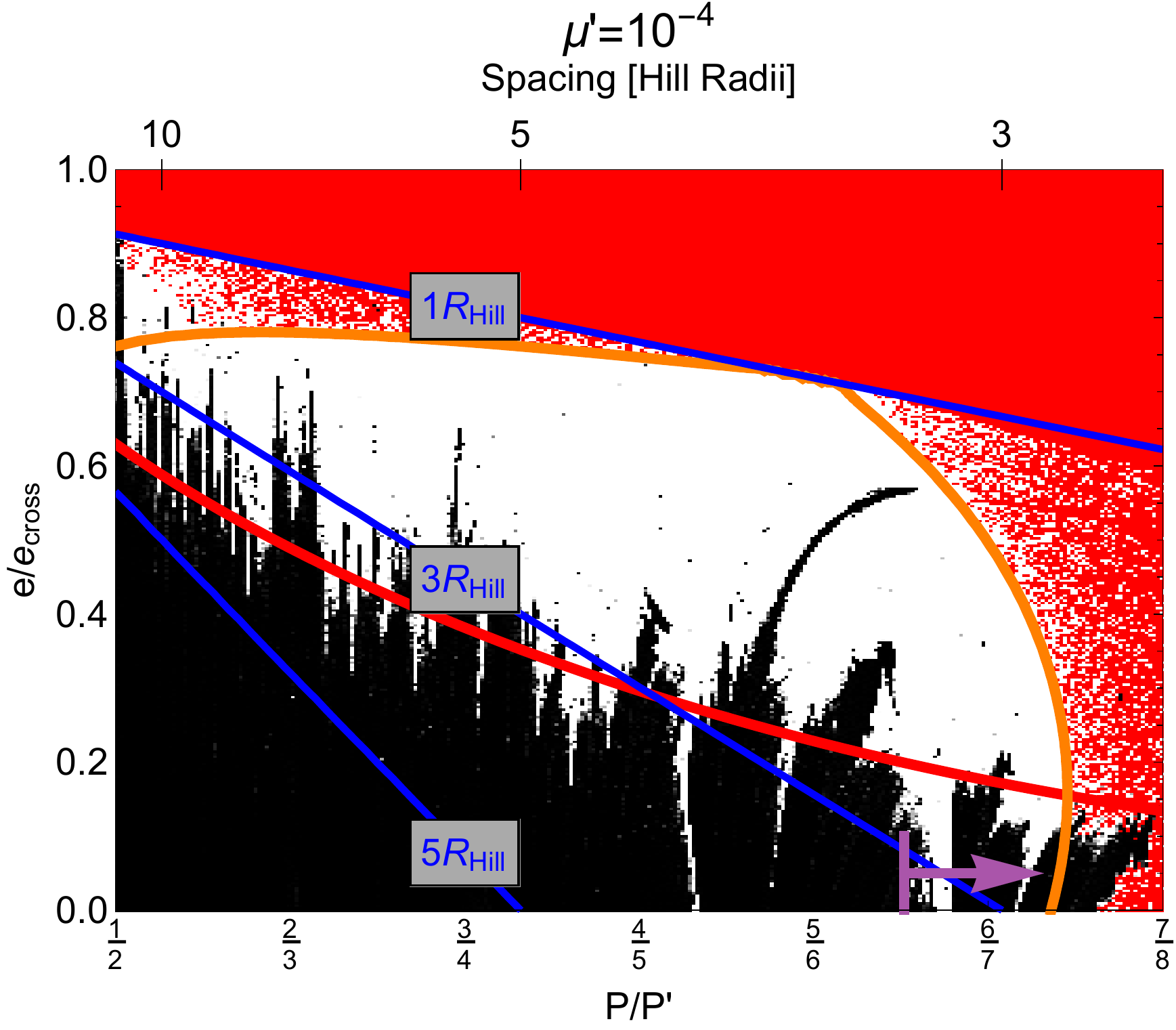}
\caption{\label{fig:criteria_compare_approach} 
Chaotic phase space structure for a test-particle subject to external perturbers of three different masses.
Lines of constant closest approach distance, $a'-a(1+e)$, in units of Hill radii, $R_\text{Hill}={a'(\mu'/3)^{1/3}}$ are shown in blue.
Various stability criteria are also plotted: {our resonance overlap boundary (Eq. \ref{eq:tau_sum}) in red, the Hill stability boundary in orange, and \citet{Wisdom80}'s resonance overlap boundary, interior to which chaos is predicted for all eccentricity values, indicated by a purple arrow.}
The semi-major axis separation, $a'-a$, is shown by ticks on the top of each panel in units of $R_\text{Hill}$. 
The background gray-scale is the same as in Figure \ref{fig:w_compare} with the addition that initial conditions leading to a close encounters (within a Hill radius) are marked in red. 
The Hill stability boundary separates initial conditions that could experience such an encounter from those that cannot based on conservation of the Jacobi constant.
}
\end{figure*}

  {
  A somewhat common practice in the literature is to presume that stability criteria derived for circular orbits can be applied to eccentric systems by simply replacing the critical semi-major axis separation, $a_2-a_1$, with the closest approach distance, $a_2(1-e_2)-a_1(1+e_1)$. 
  For example, \citet{Giuppone:2013iw} propose such a `semi-empirical' stability criterion as an  extension  \citet{Wisdom80}'s overlap criterion to eccentric planet pairs.
  Specifically, \citet{Giuppone:2013iw} posit that a pair of anti-aligned orbits will be unstable if $\frac{a_2(1-e_2)-a_1(1+e_1)}{a_2(1-e_2)}<\delta$ where $\delta=1.57[\mu_1^{2/7}+\mu_2^{2/7}]$.
  Their empirical criterion (slightly modified here to $\frac{a_2(1-e_2)-a_1(1+e_1)}{a_2(1-e_2)}<1.46(\mu_1+\mu_2)^{2/7}$ so as to match \citet{Deck2013overlap}'s prediction at $Z=0$) is plotted as a green curve in Figure \ref{fig:criteria_compare} and provides a fair approximation for the transition to chaos. 
  Figure \ref{fig:criteria_compare_approach} compares our resonance overlap prediction (red curves) to contours of constant closest approach distance for an eccentric test-particle subject to an exterior perturber for three different perturber masses. 
  The figure shows that, while our prediction matches the simulation results quite well, it cannot be reduced simply to a threshold on closest-approach distance that is independent of mass.}

{As described in the introduction, a number of empirical studies have derived relationships to predict the stability of multi-planet systems.
These empirical relations are generally derived for systems of three or more planets and cast as predictions for the timescale for instability to occur as a function of planet spacings  measured in mutual Hill radii.
Directly comparing our analytic resonance overlap criterion to these empirical studies is difficult since our analytic criterion only applies to two-planet systems and yields a binary classification of systems as chaotic or regular without any instability timescale information.
Nonetheless, we can make a couple of qualitative comparisons: first, we showed in Section \ref{sec:two_planet_overlap} that the resonance optical depth and onset of chaos depends on planet spacing measured in units of ${\mu^{1/4}a}$.
Presuming that mean-motion resonance overlap is responsible for chaos in higher-multiplicity systems,\footnote{In systems of three or more planets, overlap of secular resonances and/or three-body resonances could also play a significant role in determining dynamical stability. Since three-body resonances arise from combinations of two-body resonances their density should also depend on planet spacing measured in units of ${\mu^{1/4}a}$.
Indeed, \citet{2011MNRAS.418.1043Q} predicts that the degree of overlap of three-body resonances in three-planet systems scales with the planets' spacing measured in units of ${\mu^{1/4}a}$.} planet separation measured in units of ${\mu^{1/4}a}$ should be a better predictor of systems' stability than separations measured in Hill radii.
Second, while most of the studies mentioned in the introduction focus on circular planetary systems, \citet{PuWu2015} explore the eccentricity-dependence of stability lifetimes.
They find that more eccentric systems require slightly larger closest-approach distances in units of Hill radii to maintain the same stability lifetime as more circular systems.
This trend is consistent with the prediction of our overlap criterion shown in Figure \ref{fig:criteria_compare_approach}: more eccentric systems require  greater closest-approach distances to maintain regularity.}

 Whether or not a pair of planets is chaotic, their ultimate fate can sometimes be constrained by angular momentum and energy
 conservation laws. 
If those conservation laws forbid the pair from experiencing close encounters, the system is called {\it Hill stable}. 
In the circular restricted three-body problem, Hill stability is determined by the Jacobi constant. When a particle's Jacobi constant is greater than the Jacobi constant of a particle at the $L_1$ Lagrange point, then close encounters between the particle and perturbing mass are prohibited \citep{2000ssd..book.....M}.
{The consequences of Hill stability are evident in the distribution of initial conditions leading to close encounters in Figure \ref{fig:criteria_compare_approach}.}

A generalization of Hill stability exists for systems of three massive, gravitationally interacting bodies \citep[e.g.,][]{1982CeMec..26..311M}.
For two-planet systems with total energy $E$ and angular momentum $L$, if the product $L^2 E$ is greater than some critical value then close approaches between the planets are forbidden. 
Importantly, Hill stability does not preclude substantial changes in the planets' semi-major axes or even their ejection from the system.
\citet{1993Icar..106..247G} provides an analytic criterion for Hill stability, formulated in terms of the orbital elements of a planet pair.
The solid orange line  in Figure \ref{fig:criteria_compare} shows his result
 (from his Equation 21). 
The threshold for Hill stability is quite different from that for chaotic onset, as we discuss below.

\subsection{Long-term stability}
\label{sec:chaos_vs_stability}
\begin{figure*}
\centering
\includegraphics[width=1\textwidth]{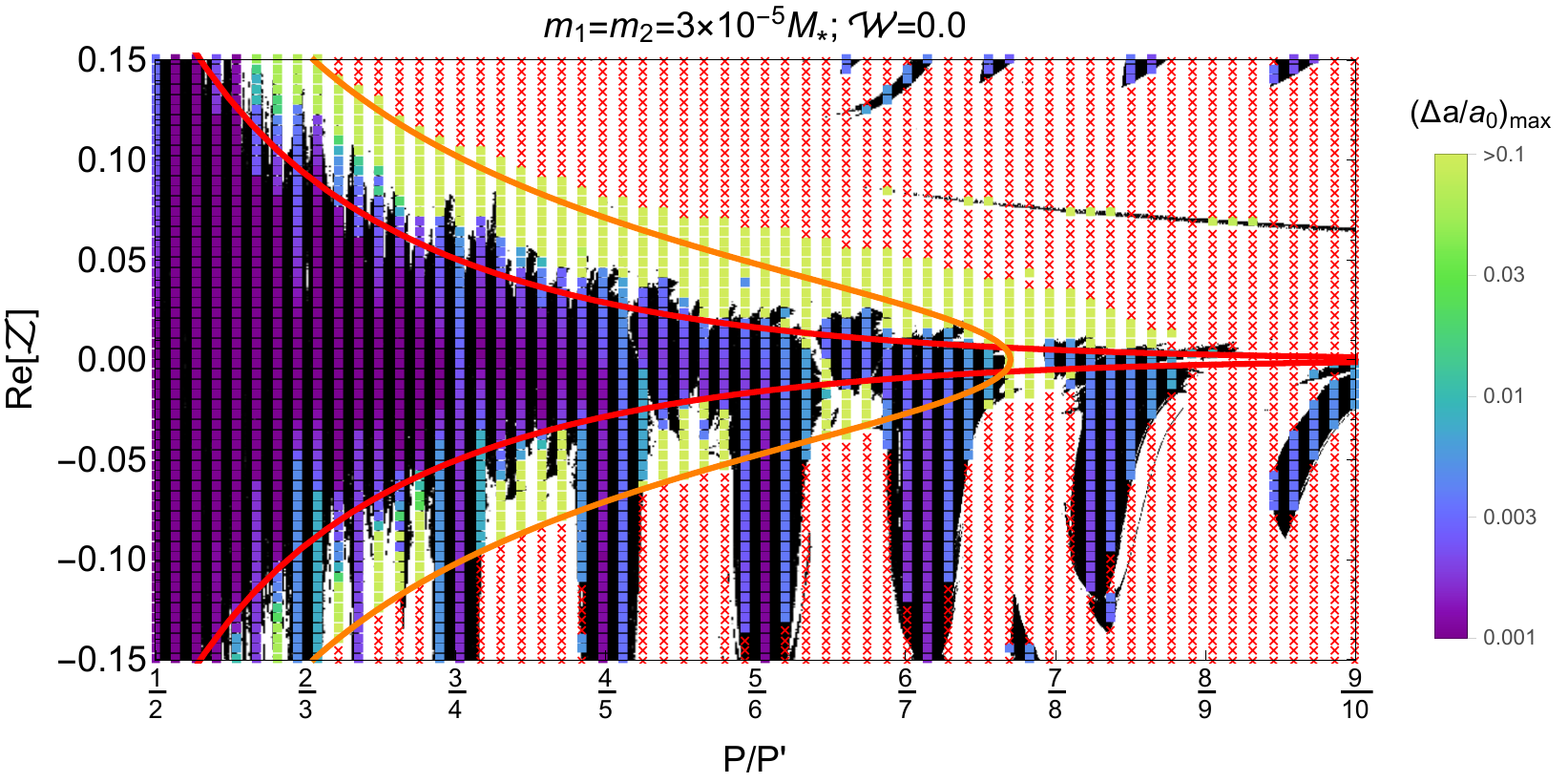}
\includegraphics[width=1\textwidth]{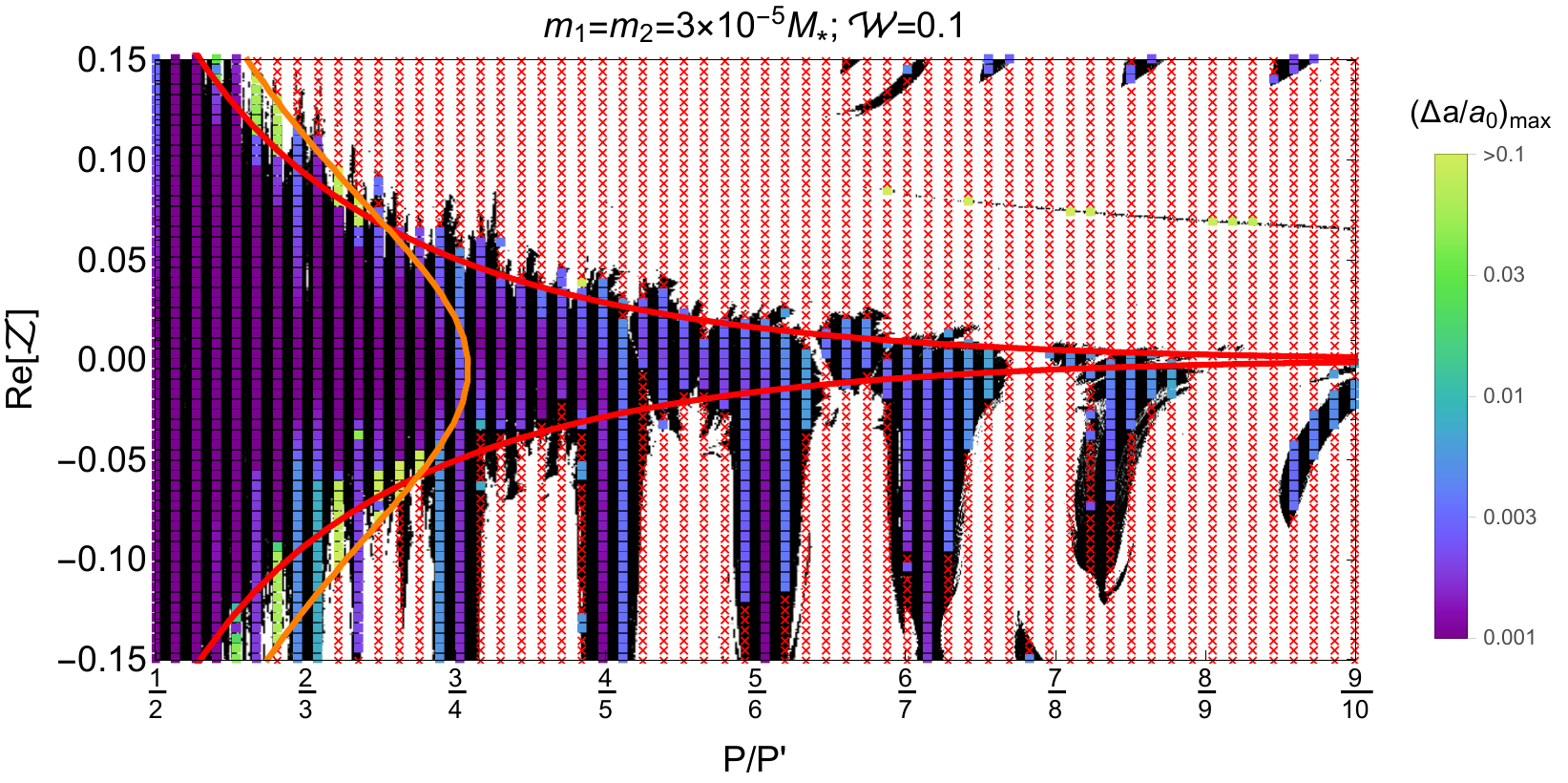}
\caption{\label{fig:stability_compare} 
Comparison between onset of chaos and long-term evolution. 
The black/white background maps are reproduced from Figure \ref{fig:w_compare}.
Red `x's and colored squares show the outcome of long-term numerical simulations that lasted
 $10^7$ orbits of the outer planet. 
 Red `x's denote integrations that were stopped early after experiencing a close encounter.  
 The purple/green color scale indicates, for systems that did not experience close encounters, the maximum fractional deviation in the inner planet's semimajor axis. 
In the top panel systems are initialized with ${\cal W}=0$ and in the bottom panel with ${\cal W}=0.1$.
Orange curves indicate the Hill stability boundary from \citet{1993Icar..106..247G}.
Our resonance overlap boundary (Equation \eqref{eq:zcrit_exact}) is plotted in red.
}
\end{figure*}

We explore the relationship between the onset of chaos and long-term stability with two suites of numerical simulations run  for $10^7$ outer planet orbits.
Figure \ref{fig:stability_compare} demonstrates
 that most systems  that do not cross the threshold for chaos (i.e., that fall within our predicted red curves) 
  exhibit little change in 
$a$ over the course of the simulation. In contrast, orbits which are chaotic can experience two different fates:  many of them experience close encounters (red `x's), but many also exhibit relatively large changes in $a$  throughout the duration
of the simulations without ever experiencing  close encounters or ejections (yellow-green squares).
We attribute the boundary between the latter two behaviors as being due to Hill stability: the yellow-green squares
are Hill stable, and the red crosses are not.
Note that this true Hill stability boundary is roughly coincident with the prediction of Gladman but there is some discrepancy, 
which is presumably due to some of Gladman's approximations.\footnote{
Specifically, Gladman's formula is given to leading order in $(\mu_1+\mu_2)^{1/3}$, neglecting terms of order $\propto \mu^{2/3}$ and higher in planet-star mass ratios. For the planets shown in Figure \ref{fig:stability_compare} this corresponds to fractional error of ${(3\times 10^{-5}})^{1/2}\approx 0.03$.
This estimated error is consistent with the percent-level deviation in period ratio between Hill-stability boundary  predicted by the orange curve in Figure \ref{fig:stability_compare} for $\ZZ=0$ ($P/P'\approx 7/8$) and the last yellow-green square near $\ZZ=0$ ($P/P'\approx 8/9$).
}
Systems such as the yellow-green squares that are chaotic yet do not experience close encounters have been referred to as ``Lagrange unstable" \citep[e.g.,][]{Deck2013overlap}. (More precisely, ``Lagrange unstable" refers to systems that experience significant semi-major axis variations, irrespective of whether or not they are Hill stable.) 
The Kepler-36 system provides an illustrative example \citep{2012Sci...337..556C,deck2012kep36}:
the two sub-Neptune planets, b and c, exhibit chaos with a Lyapunov time of only $\sim$10 years.
\citet{deck2012kep36} find that, over the course of $\sim 10^7$ years, $\sim$75\% of their integrations exhibit $>10\%$ variations in the planets' semi-major axes.
The planet pair has presumably survived for a substantially longer time, suggesting it is Hill stable and protected from close encounters.
We plot Kepler-36 b/c on Figure \ref{fig:optical_depth_2} using the planets' masses and eccentricities measured from transit timing variations in \citet{2017AJ....154....5H}.
The planet pair lies very near our prediction for the onset of chaos. 
{Finally, we note that failing the Hill criterion does not necessarily mean a planet pair is doomed to experience a close encounter; the system must be chaotic as well.
Both panels of Figure \ref{fig:stability_compare} show regions of stable initial conditions that fail the Hill criterion {(i.e., lie to the right of the orange curves)} but are protected from close encounters by first-order resonances. Additionally, the bottom panel contains a significant swath of regular points with small $Z$s that fail the Hill criterion but remain stable.
}

\section{Summary and Conclusions}
\label{sec:conclusion}
We derived a new criterion for  the threshold of instability in two-planet systems. 
The derivation was based on the idea that the onset of chaos, and therefore instability,  occurs where resonances overlap in phase space.  
Our prediction for the test-particle eccentricity at which chaos first occurs in the restricted three-body problem is given by Equation \eqref{eq:tau_sum} at $\tau_{\rm res}=1$ and is depicted as the solid red line in Figure \ref{fig:tau_compare}; Equation \eqref{eq:ecrit_approx_2} gives an adequate fitting formula for this critical eccentricity.
Our prediction for the onset of chaos generalizes from the restricted problem to two massive and eccentric planets in a straightforward manner: one simply replaces test particle's eccentricity with the planets' relative eccentricity (Eq. \ref{eq:zdiff}) and the perturber's mass with the sum of the planets' masses. This yields 
Equation \eqref{eq:zcrit_exact} with $\tau_{\rm res}=1$ as our criterion for the onset of chaos in two-planet systems with an adequate approximation given by Equation \eqref{eq:zcrit_approx}. 

This work  extends the past overlap criteria developed by \citet{Wisdom80} and \citet{Deck2013overlap} for {nearly} circular planets to eccentric planet pairs. 
The `optical depth' method adopted in Section \ref{sec:two_planet_overlap:optical_depth} allowed us to consider resonances at all orders and extend these past criteria, which treated only first-order resonances.
{Figure \ref{fig:validity_regions} shows  how our new criterion extends the range of parameters under which a two-planet system  becomes chaotic.}

The analytic overlap predictions were shown to successfully predict the onset of chaos seen in numerical simulations in Section \ref{sec:numerical_compare} {(Figures \ref{fig:chaos_compare}--\ref{fig:optical_depth_2})}.
{We also used the simulations to explore the conditions under which a chaotic system leads to planetary collisions.}

The parameter regime studied in this paper, closely spaced planets with moderate eccentricities, is motivated by the observed exoplanet population.
The results of this work serve as a starting point for better understanding the sources of chaos and instability in realistic systems.
While our overlap criterion was derived assuming strictly coplanar planets, we expect that our criterion still approximately predicts the onset of chaos when inclinations (measured in radians) are {small compared to eccentricities.
In this regime, the disturbing-function terms associated with any particular MMR will be dominated by the eccentricity-dependent terms.
Additional {development is} likely necessary to predict the onset of chaos when inclinations are comparable in size to eccentricities.}

Finally, we expect the formulae for resonance widths and the `optical depth' formulation of resonance overlap derived in this paper will prove to be useful tools for understanding the onset of chaos in more complicated systems hosting three or more planets.

\acknowledgments
Acknowledgments.
We thank {Matt Holman, Jacques Laskar, Rosemary Mardling, Alessandro Morbidelli, Matt Payne, Alice Quillen, Dan Tamayo, and Yanqin Wu  for helpful discussions and comments. We thank Jack Wisdom for his careful referee report and insightful comments.} S.H. acknowledges support from the NASA Earth and Space Science Fellowship program, grant No. NNX15AT51H. Y.L. acknowledges NSF grant AST-1352369 and NASA grant  NNX14AD21G.
 This research was supported in part through the computational resources and staff contributions provided for the Quest high performance computing facility at Northwestern University which is jointly supported by the Office of the Provost, the Office for Research, and Northwestern University Information Technology.
\software{REBOUND, python, Mathematica}
\bibliographystyle{yahapj}
\bibliography{refs}
\appendix
\section{Derivation of Disturbing Function Coefficients}
\label{sec:appendix}
\subsection{Approximation of $S_{j,k}$ for Closely spaced Planets}
\label{sec:app:res_widths}
{The $S_{j,k}$ appearing in our resonance Hamiltonian (Equation \ref{eq:ham_ctp_simple}) are defined via the following double Fourier expansion:}
\begin{eqnarray}
\frac{1}{\sqrt{r^2+a'^2-2a'r\cos(\lambda'-\theta)}}&=&\frac{1}{a'}\sum_{j=-\infty}^{\infty}\sum_{k=0}^{\infty}S_{j,k}(\alpha,e)\cos(j\psi+kM) \label{eq:dist_fn_cos}
\end{eqnarray}
where $\psi = \lambda'-\lambda$ and $M=\lambda-\varpi$ is the particle's mean anomaly.\footnote{Equation \eqref{eq:dist_fn_cos} represents a Fourier expansion of the direct part of the disturbing function. 
The Fourier expansion of the indirect piece of the the disturbing function, $a'^{-2}r\cos(\lambda'-\theta)$, only contains cosine terms $\propto\cos(\psi+kM)$. These Fourier harmonics, which have $j=1$ in Equation \eqref{eq:dist_fn_cos}, do not contribute to the MMRs considered in this paper and therefore we ignore the indirect piece of the disturbing function.
}
{Note that the argument of the cosine at a given $j$ and $k$ is $j(\lambda'-\lambda)+k(\lambda-\varpi)$, which is the same as in Eq. \ref{eq:ham_ctp_simple} after setting $\lambda'=t(j-k)/j$.}
Our goal is to determine an expression that can be used to compute $S_{j,k}$.
We compare the above expression with the Fourier expansion in terms 
of Laplace coefficients, $b_{1/2}^{(j)}$ :
\begin{eqnarray}
\frac{1}{\sqrt{r^2+a'^2-2a'r\cos(\lambda'-\theta)}}&=&\frac{1}{a'}\sum_{j=-\infty}^{\infty}\frac{1}{4}b_{1/2}^{(j)}\left(\frac{r}{a'}\right)e^{{ij(\lambda'-\theta)}} + c.c. \label{eq:laplace_expansion}\\
 &=&{1\over a'}\sum_{j=-\infty}^{\infty}\frac{1}{4}
b_{1/2}^{(j)}\left(\alpha\frac{r}{a}\right)e^{ij(\lambda-\theta)}e^{ij\psi} + c.c.
\label{eq:potential} 
\end{eqnarray}
\citep{2000ssd..book.....M}. Since  $r/a$ and $\lambda-\theta$ in the latter expression depend only on $e$ and $M$, 
we deduce that 
\begin{eqnarray}
S_{j,k}(\alpha,e) &=& 
{1\over 4\pi }\int_0^{2\pi}b_{1/2}^{(j)}\left(\alpha{r\over a}  \right)\exp[{ij(\lambda-\theta)-ikM}] dM+c.c. \label{eq:Sjk_with_M}
\end{eqnarray}
The coefficients $S_{j,k}(\alpha,e)$ can be computed exactly via numerical integration of Equation \eqref{eq:Sjk_with_M}.\footnote{
To make the integrand an explicit function of the integration variable, we introduce the eccentric anomaly {$u$} to write  
$r=a(1-e\cos u)$, $M=u-e\sin u$, $\cos(\theta-\varpi)=\frac{\cos u -e}{1-e\cos u}$  and  $\sin(\theta-\varpi)=\frac{\sqrt{1-e^2}\sin u}{1-e\cos u}$ \citep[e.g.,][]{2000ssd..book.....M}, {which yields}
\begin{equation*}
S_{j,k}(\alpha,e)  ={1\over 4\pi }\int_0^{2\pi}b_{1/2}^{(j)}\left[\alpha(1-e\cos u) \right]\fracbrac{1-e\cos u}{\cos u - e +i\sqrt{1-e^2}\sin u}^{j}\exp[{i(j-k)(u-e\sin u )}](1-e\cos u)du + c.c. \label{eq:Sjk_with_u}
\end{equation*}
}

We will now derive an approximation in the limit of closely spaced planets.
For a particle near a $j$:$j-k$ resonance in this limit we have $\alpha\approx 1- \frac{2k}{3j}$.
We define $y=e/\eX \approx \frac{3j}{2k}e$ so that 
\begin{eqnarray}
r &\approx& 1-\frac{2k}{3j}y\cos M + {\cal O}\fracbrac{2ky}{3j}^2\label{eq:r_approx}\\
\theta &\approx& \lambda-\frac{4k}{3j}y\sin M + {\cal O}\fracbrac{2ky}{3j}^2\label{eq:theta_approx}
\end{eqnarray}.
Inserting Equations \eqref{eq:r_approx} and \eqref{eq:theta_approx} into Equation \eqref{eq:Sjk_with_M} and using the approximation 
\begin{equation*} b^{(j)}_{1/2}(\alpha)\approx (2/\pi)K_0(j|1-\alpha|)
\end{equation*}
for the Laplace coefficients,  where $K_0$ is a modified Bessel function of the second kind \citep{1981ApJ...243.1062G}, we have
\begin{eqnarray}
S_{j,k}(\alpha,e) &\approx & s_{k}(y)\equiv \frac{1}{\pi^2}\int_0^{2\pi}K_0\left[\frac{2k}{3}(1+y\cos M)\right]\cos\left[k\left(M+\frac{4}{3} y\sin M\right)\right] dM~.\label{eq:Sk_def}
\end{eqnarray}

\subsection{$s_k(y)$ to Leading Order in Eccentricity}
\label{sec:app:leading_order}
We now derive the leading order term in $y$  for $s_{k}$ as follows:
first, we define $\xi = ye^{iM}$ and its complex conjugate, $\xi^*=ye^{-iM}$, as well as 
\begin{eqnarray}
I_k(\xi,\xi^*)\equiv K_0\left[\frac{2k}{3}\left(1+\frac{\xi +\xi^*}{2}\right)\right]\exp\left(\frac{2k}{3}(\xi^*-\xi)\right)~.
\end{eqnarray}
We may then rewrite Equation \eqref{eq:Sk_def} as
\begin{eqnarray}
s_{k}(y)&=& \frac{1}{2\pi^2}\int_0^{2\pi}e^{-ikM}I_k(\xi,\xi^*)dM + c.c.\nonumber\\
&=&\frac{1}{2\pi^2}\int_{0}^{2\pi} e^{-ikM}\left(\sum_{k'=0}^\infty\frac{1}{k'!}\frac{d^{k'}I_k(\xi,\xi^*)}{d\xi^{k'}}\bigg|_{\xi=0}\xi^{k'} \right)dM + c.c \nonumber \\
&=& \frac{2}{\pi}\left(\frac{1}{k!}\frac{d^kI_k(\xi,0)}{d\xi^k}\bigg|_{\xi=0}+ {\cal O}{(|\xi|^{2})}\right)y^{k}\label{EQ:APP:SJKINT_EXP}\nonumber\\
&=& \frac{2}{\pi}\frac{1}{k!}\frac{d}{dx}\left\{K_0\left[\frac{2k}{3}\left(1+\frac{x}{2}\right)\right]\exp\left(-\frac{2k}{3}x\right)\right\}_{x=0}y^k+{\cal O}(y^{k+2})~.
\end{eqnarray}
In the large-$k$ limit we approximate $K_{0}(x)\approx\sqrt[]{\frac{\pi}{2x}}\exp({-x})$, in which case 
\begin{eqnarray}
s_{k}(y)
&\approx&\frac{2}{\pi} \frac{1}{k!} 
	\sqrt{\frac{3\pi}{4k}}\exp({-2k/3})
	\frac{d^k}{dx^k}
	\left[
		(1+x/2)^{-1/2}\exp(-{kx})
    \right]_{\delta=0}y^k \nonumber
\\
&=&\frac{\sqrt{3} \exp({-2k/3})}{\sqrt{\pi k}} 
	(-1)^k\frac{k^k}{k!}y^k\sum_{m=0}^{k}\binom{-1/2}{m}\left(-\frac{1}{2k}\right)^m\frac{k!}{(k-m)!}\label{eq:wk_approx1} \\
&\approx&\frac{(-1)^k\sqrt{3} \exp({k/3})}{\pi k}y^k \label{eq:wk_approx2}
\end{eqnarray}
where we have used Stirling's approximation $\frac{k^k}{k!}\approx e^k/\sqrt[]{2\pi k}$ and replaced the sum in Equation \eqref{eq:wk_approx1} 
with its large-$k$ limit, $\sqrt{2}$, to derive Equation \eqref{eq:wk_approx2}.
\footnote{\citet{Holman:1996tu} derive an approximation for the disturbing function coefficients of high-order resonances similar to Equation \eqref{eq:wk_approx2}. We find that introducing the complex variable $\xi$ greatly simplifies their original derivation.  While \citet{Holman:1996tu}'s formula similarly predicts $S_{j,k}\propto \fracbrac{e}{\eX}^k$, their final expression contains a slightly different numerical pre-factor than Equation \eqref{eq:wk_approx2}. We find good agreement when comparing our approximation, Equation \eqref{eq:wk_approx2}, to values of $S_{j,k}$ computed exactly via Equation \eqref{eq:Sjk_with_M} indicating \citet{Holman:1996tu}'s derived expression contains an error.}
\subsection{Single-harmonic Approximation Versus Exact Resonance Widths}
\label{sec:appendix:pendulum_vs_exact}
\begin{figure}
    \centering
    \includegraphics{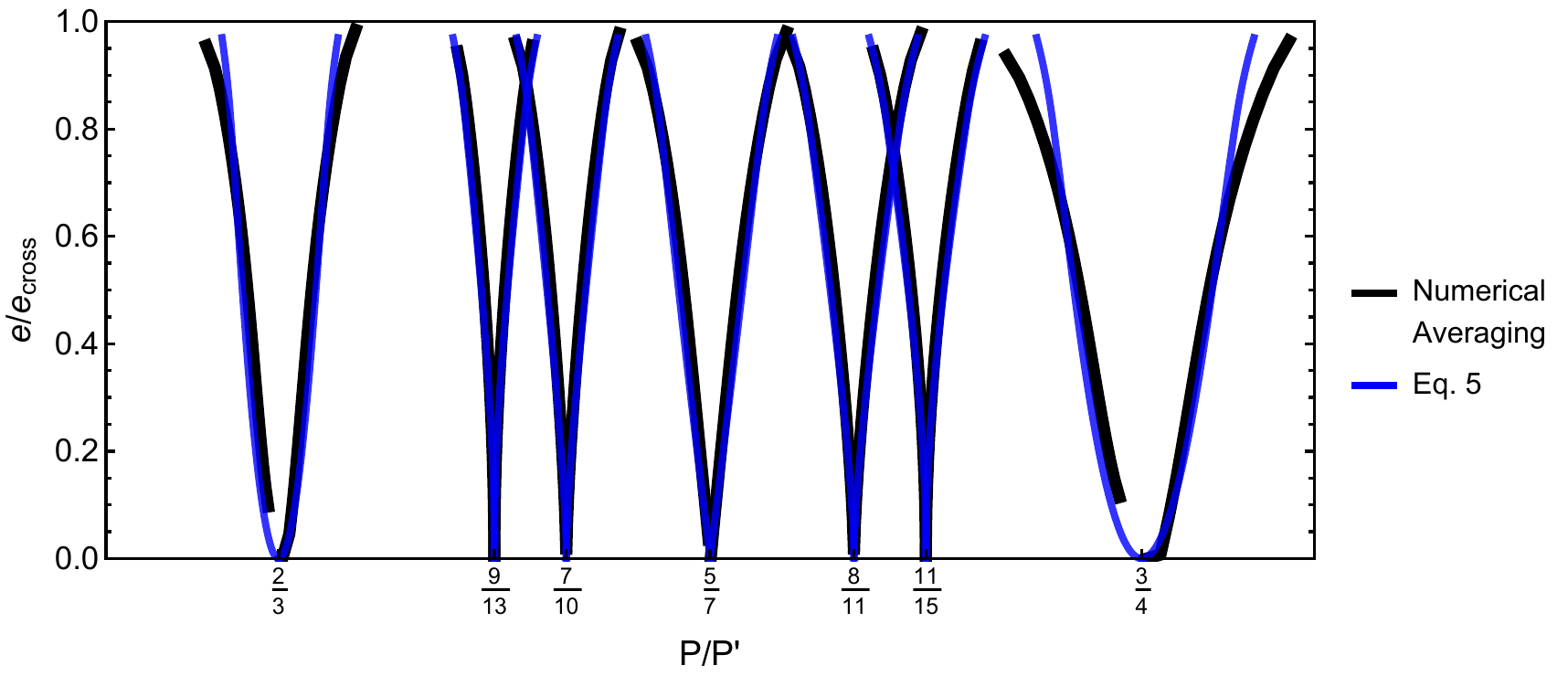}
    \caption{A comparison between resonance widths computed via the single-harmonic approximation versus numerical averaging for resonances up to fourth order between the 3:2 and 4:3 MMRs.
    Blue lines show single-harmonic model resonance separatrices computed using Equation \eqref{eq:da_ctp} (with $S_{j,k}$ given by Equation \eqref{eq:Sjk_with_M}).
    Black lines show resonance separatrices computed via numerical averaging following the methods of \citet{Morbidelli:1995jf}. 
    Equation \eqref{eq:da_ctp} does not capture the special behavior of first-order resonances near $e=0$, where the separatrix wide of the nominal resonance location disappears as seen in resonance widths computed via numerical averaging.
    }
    \label{fig:res_width_method_compare}
\end{figure}

 The Hamiltonian model used in Section \ref{sec:two_planet_overlap:widths} to derive the width of a $j$:$j-k$ MMR included only the single cosine term, $S_{j,k}(\alpha,e)\cos(\phi)$, where $\phi = j\psi + k M$, from the Fourier expansion in Eq. \eqref{eq:dist_fn_cos}.  
A proper treatment of the resonant dynamics should really include all terms of the form $S_{nj,nk}(\alpha,e)\cos(n\phi)$ with $n=1,2,...$ from the Fourier expansion.
The sum of all these terms is simply the Fourier representation of the averaged disturbing function, i.e.,
\begin{eqnarray}
\sum_{n=1}^{\infty}S_{nj,nk}(\alpha,e)\cos(n\phi)=\bar{R}(\phi;\alpha,e) \equiv \frac{1}{2\pi}\int_{0}^{2\pi}\frac{dM}{\sqrt{r^2+a'^2-2a'r\cos(\lambda'-\theta)}}\bigg|_{j\psi+k M=\phi}~,\label{eq:avg_distfn}
\end{eqnarray}
which can be computed numerically by setting $\lambda'=\phi-\frac{k}{j}M$ and evaluating $r$ and $\theta$ as functions of $M$ in the integrand.
We follow the methods of \citet{Morbidelli:1995jf} to compute resonance widths using the averaged disturbing function $\bar{R}(\phi;\alpha,e)$.
Briefly, this is done as follows: first we  reduce the Hamiltonian via a canonical transformation to a single degree-of-freedom system that depends on $\phi$ and its conjugate momentum, $I$, as well as a conserved quantity, $K$, as in Section \ref{sec:two_planet_overlap:widths}.
Next, for a given $K$, we identify the Hamiltonian contour in the $I-\phi$ plane corresponding to the separatrix and determine its maximum width.
Finally, the pair of points on the separatrix at its maximum width are converted to a pair of points in the $a$-$e$ plane.
Repeating this procedure over a range of $K$s allows us to  construct resonance widths in the $a$-$e$ plane.
In Figure \ref{fig:res_width_method_compare}  we compare numerically computed resonance widths with the prediction of the simplified ``single-harmonic" model.
The agreement between the single-harmonic model and numerically computed resonance widths is quite good, justifying our use of the single-harmonic model resonance widths to calculate the resonance optical depth.

    {For completeness, we mention that \citet{Ferraz-Mello_Sato_1989} also develop numerical techniques for computing Hamiltonian models of resonant motion at high eccentricities. Their method is similar to \citet{Morbidelli:1995jf}'s method described above, except that, before conducting a numerical average, they first expand the un-averaged disturbing function in a double Taylor series in the variables $e\cos\phi$ and $e\sin\phi$ about a specified libration center.}

\section{Comparison of Integration Methods}
\label{sec:appendix:integrator_compare}
\begin{figure}
    \centering
    \includegraphics[width=0.8\textwidth]{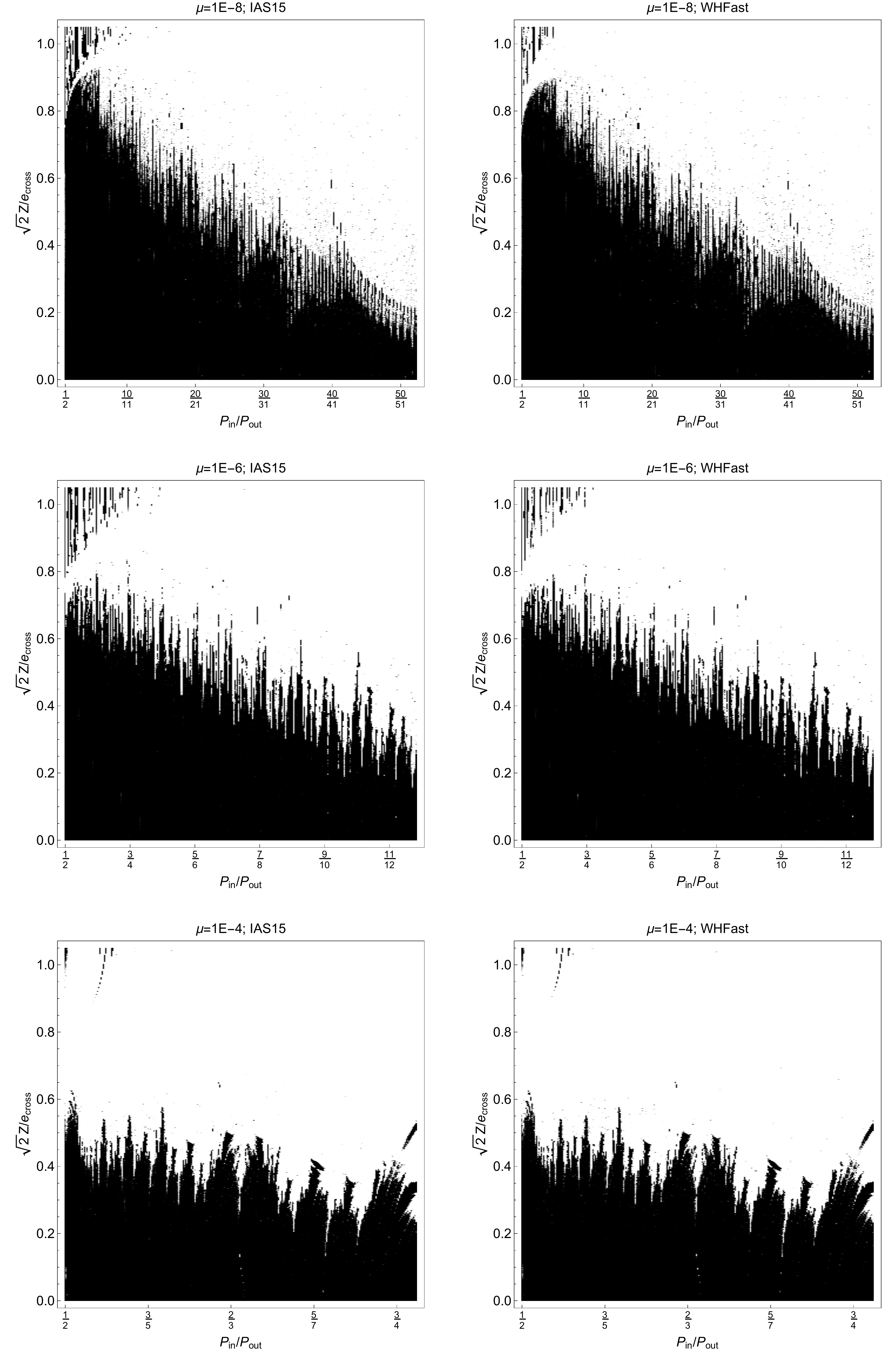}
    \caption{
    A comparison between MEGNO maps computed with the IAS15 integration scheme, shown in the left column, and the symplectic WHFast, shown in the right column. A different planet-star mass ratios, $\mu$, is shown in each row. The color scheme is the same as in Figure \ref{fig:w_compare}, with chaotic trajectories in white and regular trajectories shown in black. Initial conditions and details of the numerical methods are described in the text.}
    \label{fig:compare_methods}
\end{figure}

{
    Here we explore whether numerical artifacts influence the results in Section \ref{sec:numerical_compare} obtained with the Wisdom-Holman integrator, WHFast implemented by \citet{RTwhfast2015}. Overlapping  resonances between the integration time step and a planet pairs' synodic period can cause spurious chaotic behavior in integrations using the Wisdom-Holman algorithm, especially for closely spaced planets like those considered here \citep{WisdomHolman1992stepsize}.  Eccentric orbits can also lead to artificial chaos if the time step is not small enough resolve perihelion passage \citep{RauchHolman1999,wisdom2015}.
    \citet{wisdom2015} finds that a step size of 1/20 of the perihelion passage timescale, defined as $T_{p}=2\pi/\dot{f}_p$ where $\dot{f}_p$ is the rate of change of the true anomaly at pericenter, is typically adequate to resolve perihelion passage. Our choice of time step, 1/30th the inner planet's orbital period, provides 20 or more steps for eccentricities up to $e=0.2$ or, equivalently, a combined eccentricity $Z\approx \sqrt{2}e = 0.3$ (assuming ${\cal W}=0$). Thus, our time step should be adequate for most of the parameter space considered in Section \ref{sec:numerical_compare}.
    To ensure that our results are not driven by artificial chaos induced by our numerical method we compare the results obtained with the WHFast integrator  with the high-order, adaptive-timestep IAS15 integrator \citep{RS15}. Figure \ref{fig:compare_methods}  shows the result of this comparison. Each panel shows a MEGNO map in period ratio versus $\sqrt{2}Z/\eX$ for a pair of equal-mass planets. Integrations are run for 3000 orbits of the inner planet or until a close encounter within one mutual Hill radius,  $a_2(2\mu/3)^{1/3}$, occurs. Planets are initialize with $\lambda_1=\lambda_2=0$, and eccentricities and longitudes of perihelia such that $\arg({\cal Z})=0$ and ${\cal W}=0$.    All WHFast integrations are conducted with a time step of 1/30th the inner planet's initial orbital period. All IAS15 integrations are conducted with the precision parameter, $\epsilon_b$, set to its default value,  $\epsilon_b=10^{-9}$ \citep[see][]{RS15}. The integration methods show good agreement across the broad range of planet masses, eccentricities, and spacings plotted in Figure \ref{fig:compare_methods}.}
\listofchanges
\end{document}